\documentclass[preprint]{elsarticle}

\usepackage{amssymb}
\usepackage{amsmath}
\usepackage{lineno}
\usepackage[utf8]{inputenc} 
\usepackage{graphicx} 
\usepackage[colorlinks=true]{hyperref}
\usepackage{cleveref}

\usepackage{tabularx}
\usepackage{float}
\usepackage{threeparttable}
\usepackage{multirow}
\usepackage[utf8]{inputenc} 
\usepackage{graphicx} 
\usepackage{comment}
\usepackage{enumitem}
\usepackage{booktabs}
\usepackage[left=1in, right=1in]{geometry}
\newlist{tabitemize}{itemize}{1}
\setlist[tabitemize]{nosep,
                  topsep= 0pt,
                  partopsep=0pt,
                  leftmargin= *,
                  label=\textbullet,
                  before=\vspace{0.3\baselineskip},
                  after=\vspace{-\baselineskip}
                  }

\newcolumntype{M}[1]{>{\centering\arraybackslash}m{#1}}
\newcolumntype{Y}{>{\centering\arraybackslash}X}
\graphicspath{{./figures/}}

\journal{xxx}

\begin{document}

\begin{frontmatter}

\title{Hybrid modeling approach for better identification of building thermal network model and improved prediction}

\author[inst1]{Sang woo Ham}
\author[inst1]{Donghun Kim}
\affiliation[inst1]{organization={Building Technology \& Urban Systems Division},
            addressline={Lawrence Berkeley National Laboratory}, 
            city={Berkeley}, 
            state={CA},
            country={USA}}

\begin{abstract}

The gray-box modeling approach, which uses a semi-physical thermal network model, has been widely used in building prediction applications, such as model predictive control (MPC). However, unmeasured disturbances, such as occupants, lighting, and in/exfiltration loads, make it challenging to apply this approach to practical buildings. In this study, we propose a hybrid modeling approach that integrates the gray-box model with a model for unmeasured disturbance. After reviewing several system identification approaches, we systematically designed the unmeasured disturbance model with a model selection process based on statistical tests to make it robust. We generated data based on the building model calibrated by real operational data and then trained the hybrid model for two different weather conditions. The Hybrid model approach demonstrates the reduction of RMSE approximately 0.2-0.9$^\circ$C and 0.3-2$^\circ$C on 1-day ahead temperature prediction compared to the Conventional approach for mild (Berkeley, CA) and cold (Chicago, IL) climates, respectively. In addition, this approach was applied for experimental data obtained from the laboratory building to be used for the MPC application, showing superior prediction performances.

\end{abstract}


\begin{highlights}

\item Development of a hybrid modeling approach that enhances the long-term temperature or load predictions of a gray-box model by combining a machine-learning model for unmeasured disturbances.

\item Investigation of the limitations of various system identification approaches for a gray-box model under unmeasured disturbances.

\item Development of the design and model selection processes for the robust hybrid modeling approach, considering the building thermal process.

\item Achieved better long-term (1-day) prediction performance of the hybrid approach compared to the gray-box model in advanced predictive control applications for both simulated and experimental data.

\end{highlights}

\begin{keyword}
Gray-box model \sep Building modeling \sep Unmeasured disturbances \sep Hybrid modeling \sep Neural network \sep Building control \sep Machine learning
\end{keyword}

\end{frontmatter}

\section*{Nomenclature}
\begin{description}

\item[CONV:] Conventional system identification
\item[HVAC:] Heating, ventilation, and air-conditioning
\item[HP-RTU:] Heat pump rooftop unit
\item[ID:] Input disturbance system identification
\item[MPC:] Model predictive control
\item[OD:] Output disturbance system identification
\item[RTF:] Runtime fraction
\item[RTU:] Rooftop unit
\item[RMSE:] Root mean squared error
\item[LD:] Lumped term for all the unmeasured disturbances

\item[($\textbf{A}(\cdot),\,\textbf{B}_{\text{u}}(\cdot),\,\textbf{B}_{\text{w}}(\cdot),\,\textbf{B}_{\text{g}}(\cdot),\,\textbf{C}(\cdot)$):] A state space model structure that maps $\theta$ to building dynamics (i.e., $G_{\text{u}}$, $G_{\text{w}}$, and $G_{\text{g}}$)
\item[($\textbf{A}_\text{d}(\cdot),\,\textbf{B}_{\text{d,u}}(\cdot),\,\textbf{B}_{\text{d,w}}(\cdot),\,\textbf{B}_{\text{g,w}}(\cdot),\,\textbf{C}_\text{d}(\cdot)$):] A discretized state space model of ($\textbf{A}(\cdot),\,\textbf{B}_{\text{u}}(\cdot),\,\textbf{B}_{\text{w}}(\cdot),\,\textbf{B}_{\text{g}}(\cdot),\,\textbf{C}(\cdot))$

\item[$A_{\text{win}}$:] Effective window area of a zone window [$\text{kW/m} ^2$]
\item[$\textbf{b}$:] Bias vector of neural network.
\item[$C_{\text{w},i}$:] Thermal capacitance of wall mass of $i$th zone [$\text{kWh/K}$]
\item[$C_{\text{za},i}$:] Thermal capacitance of zone air of $i$th zone [$\text{kWh/K}$]
\item[$\textbf{c}$:] Cell state vector in LSTM
\item[dow:] Day of week [-]
\item[$e$:] Zero mean white noise
\item[($\mathcal{F}(\cdot),\,\mathcal{G}(\cdot)$):] A state space model structure that maps $\rho$ to lumped disturbance dynamics (i.e., $H$)
\item[$f$:] Convective fraction of the incident solar radiation of a zone window [-]
\item[$G_\text{u}$:] A dynamic system that maps $\textbf{u}$ to $y_{\text{za}}$
\item[$G_\text{w}$:] A dynamic system that maps $\textbf{w}$ to $y_{\text{za}}$
\item[$G_\text{g}$:] A dynamic system that maps $\dot{Q}_{g}$ to $y_{\text{za}}$
\item[how:] Hour of week [h]
\item[hod:] Hour of day [h]
\item[$H$:] Dynamics of lumped output disturbances
\item[$\mathbf{h}$:] Hidden state vector in RNN/LSTM
\item[$i_{\text{heat}}, i_{\text{cool}}$:] Binary indicators for heating/cooling stage [-]
\item[$n_{*}$:] Number of *.
\item[$n_{\psi}$: Number of input features in disturbance model [-]
\item[$n_{\text{k},\psi}$:] Input sequence length for disturbance model [-]
\item[$n_{\text{k},\xi}$:] Output sequence length for disturbance model [-]
\item[$n_\text{layer}$:] Number of hidden layers in neural network [-]
\item[$n_\text{z}$:] Size of hidden layer [-]
\item[$n_\text{channel}$:] Number of convolution channels [-]
\item[$n_\text{filter}$:] Size of convolution filter [-]
\item[$n_\text{pool}$:] Pooling size in CNN [-]

\item[$\dot{Q}_{\text{g}}$:] Unmeasured heat gains of a zone [kW]
\item[$\dot{q}_{\text{sol,win}}$:] Incident solar radiation per area on a zone window [$\text{kW/m} ^2$]
\item[$\dot{Q}_{\text{hc}}$:] Rated heating($\dot{Q}_{\text{h}}$)/cooling($\dot{Q}_{\text{c}}$) capacity of a zone HVAC unit [kW]
\item[($R_{\text{zw}},\,R_{\text{zo}}$):] Thermal resistances between temperature nodes of a zone [$\text{K/kW}$]
\item[$t,\,k$:] Continuous and discrete time
%\item[($T_\text{l},T_\text{u}$):] Lower and upper temperature bounds [$^\circ$C]
\item[$T_{\text{za}}$:] Air temperature of a zone [$^\circ$C]
\item[$T_{\text{w}}$:] Wall thermal mass temperature of a zone [$^\circ$C]
\item[$T_\text{oa}$:] Outdoor air temperature [$^\circ$C]
\item[$T_{\text{csp}}$:] Cooling setpoint temperature [$^\circ$C]
\item[$T_{\text{hsp}}$:] Heating setpoint temperature [$^\circ$C]
\item[$T_s$:] Sampling time [s]
\item[$\textbf{u}$:] Vector of control inputs (i.e., heating or cooling operation stages, [$u_{\text{h}}$, $u_{\text{c}}$])
\item[$u_{\text{h}},\,u_{\text{c}}$:] Heating and cooling stages of a HVAC unit [-].
\item[$u_{\text{hc}}$:] Combined heating/cooling control signal [-]

\item[$\textbf{W}$:] Weight matrix of neural network.
\item[$\textbf{w}$:] Vector of measured disturbances (i.e., [$T_{\text{oa}}$, $\dot{Q}_{\text{sol,win}}$])
\item[weekday:] Binary indicator of weekday/weekend [-]
\item[$\textbf{x}$:] Vector of state variables (i.e., [$T_{\text{w}}$, $T_{\text{za}}$])
\item[$\hat{\textbf{x}}(k|j)$:] Vector of estimated (predicted) state variables at time $k$ from the data at $j$
\item[$\boldsymbol{\nu}$:] Vector of lumped output disturbances [$^\circ$C]
\item[$y_\text{za}$:] Measured thermostat temperature of a zone [$^\circ$C]

\item[$\beta_{*}$]: Regression parameters of * variable in model selection process.

\item[$\varepsilon$:] One step ahead prediction error
\item[$\boldsymbol{\Sigma}_{\textbf{x}_\text{ID}}$:] State noise covariance matrix
\item[$\boldsymbol{\Sigma}_{y_\text{za}}$:] Measurement noise covariance matrix
\item[$\boldsymbol{\zeta}$:] Vector of internal state of lumped output disturbances
\item[$\theta$:] Physical parameters consisting of thermal resistances and capacitances, [$C_{\text{w}}$, $C_{\text{za}}$, $R_{\text{zw}}$,  $R_{\text{zo}}$, $f$,$A_{\text{win}}$, $\dot{Q}_{\text{h}}$, $\dot{Q}_{\text{c}}$]
\item[$\rho$:] Parameters that constructs dynamics of lumped output disturbances, i.e. $H$
\item[($\omega_\text{l},\omega_\text{u}$):] Weights on optimization variables for ($\Gamma_\text{l},\,\Gamma_\text{u}$)
%\item[$\omega_{d}$:] Weight on optimization variables for $\delta$
\item[$\mathcal{D}(k)$:] Set of measured data from time step from timestep from 1 to $k$. 
\item[$\chi_{*}$]: Independent variable (*) of the regression in model selection process.
\item[$\upsilon_{*}(c)$:] Temperature prediction error by using unmeasured disturbance model on * dataset (i.e., either train or test). 
\item[$\boldsymbol{\psi}$:] Input vector of unmeasured disturbance model.
\item[$\varphi$:] Activation function in neural network
\item[$\boldsymbol{\xi}$:] Output vector of unmeasured disturbance model.

\end{description}

%\begin{linenumbers}

\section{Introduction}
\label{sec:introduction}

The gray-box modeling approach, often referred to as the semi-physical thermal network model, is extensively utilized in building energy prediction applications. Its applications include model predictive control (MPC) \cite{De_Coninck2016-oa,Kim2020-bo} for enhancing energy efficiency \cite{Blum2022-yq} and advancing decarbonization efforts \cite{Kim2022-tm}, due to its flexible structure. This method models a complex building using a simplified R-C (thermal resistance and capacitance) network model, while retaining the physical principles of the building’s thermal characteristics.

The conventional system identification method for gray-box models generally involves estimating parameters by minimizing the sum of squared errors between the predicted and actual temperatures $n$ steps ahead \cite{Braun2002-iv,Li2015-xm}. This technique uses readily accessible data, such as disturbances (weather conditions) and control signals (heating and cooling operation signals). However, there are also disturbances that are not easily measurable like internal heat gains from occupants, lighting, appliances, and air infiltration or exfiltration. These sources can contribute significantly to the overall heat gain but are challenging to quantify in actual buildings due to the lack of sensors, which can degrade the quality of system identification outcomes \cite{Kim2016-sysid,Kim2018-sysid}. In literature, the aggregate of heat gains from unmeasured disturbance is often simplified into a single term, presumed to be proportional to the total of all electrical loads \cite{Joe2017-cj,Blum2022-yq}. Nonetheless, installing the necessary power sensors for accurate measurement is expensive and typically unfeasible, especially in small and medium commercial buildings.

The failure in system identification can lead to significant issues in predicting building energy performance. For instance, unmeasured disturbances—such as all unaccounted internal heat gains—might be expressed as exaggeratedly high thermal capacitances and solar heat gains in the system identification results. Consequently, this misrepresentation could worsen the system’s ability to accurately predict the building’s thermal behavior in response to future heating and cooling control signals.

Various system identification techniques have been proposed to accurately estimate the model parameters of a gray-box model under unmeasured disturbances. Kim et al. \cite{Kim2016-sysid} proposed a system identification algorithm that accounts for the dynamics of unmeasured disturbances. They treated all the internal heat gains as a lumped unmeasured disturbance term (LD) and appended its dynamics to the building gray-box model for system identification. The authors utilized the impact of LD on the measurement (i.e., zone air temperature), referred to as the output disturbance (OD) approach, and solved the system identification problem using the prediction-error method (i.e., minimizing one-step prediction errors with state filtering) \cite{Ljung1998-nt}. Coffman and Barooah \cite{Coffman2018-nq} suggested a similar but slightly different approach by directly including the LD term as a heat gain in the gray-box model (i.e., the input disturbance (ID) approach). They achieved system identification by minimizing one-step-ahead prediction errors via the Kalman filter, but the selection of state variances served as tuning parameters to correctly capture the variations of the IDs. Zeng et al. \cite{Zeng2021-ok} further developed the ID approach in the frequency domain with physical constraints. Although this approach makes the system identification problem a convex optimization problem, the parameter values of the gray-box model may lose their physical scales. On the other hand, the LD term has also been modeled as a black-box function of time-related features (e.g., time of day, day of the week, and day of the year) via a feed-forward neural network, and the system identification was conducted together with the neural network \cite{Ellis2021-om}. Similarly, Kumar et al. \cite{Kumar2023-lh} modeled the LD term as piecewise constants and included it in the system identification. Having a separate model for the LD term in the system identification can capture the complex and non-linear characteristics of unmeasured disturbances, but an iterative process may be required to prevent the overall building thermal dynamics from being overfitted by the model. Finally, the theoretical analysis of the LD approach \cite{Kim2018-sysid} claims that the success of system identification under unmeasured disturbances heavily relies on the quality of training data (i.e., data with uncorrelated control inputs, and measured and unmeasured disturbances). This can be achieved by manually assigning cooling or heating signals (exciting the system) for sufficient periods \cite{Madsen1993-wf}. Nevertheless, in practice, extensive manual excitation is often not feasible for buildings in use, given the slow thermal response of the building and the seasonal characteristics of outdoor conditions.

The predictive accuracy of the gray-box model for future scenarios is often compromised, even with precisely identified system parameters, if it does not account for unmeasured disturbance profiles. In building simulations, such disturbances—which include heat gains from occupants, appliances, and infiltration—are typically represented by scheduled values, either deterministic \cite{Ashrae2017-gw} or stochastic \cite{Wilson2022-zi}. While this method is appropriate for white-box simulations aimed at estimating average energy usage based on standard disturbance schedules for building design, it may not suffice for precise predictive applications for real buildings.

Therefore, estimating unmeasured disturbances from available data is essential. Although a Kalman filter \cite{ONeill2010-fj,Coffman2018-nq} or a particle filter \cite{Ham2021-jp} can derive these disturbances from measured data, they do not offer future predictions. To address this, some research \cite{Dong2016-oq,Lee2021-ee} has developed black-box models that forecast non-HVAC energy consumption based on historical data. Additionally, two studies \cite{Ellis2021-om,Kumar2023-lh} have suggested employing a neural network-based black-box model to detect unmeasured disturbances using temporal features and incorporate them into MPC. One study conducted simultaneous identification of both the gray-box and neural network models, while the other trained the neural network model subsequent to the gray-box model identification. Yet, the concern regarding the potential overfitting of neural network models remains unresolved. Additionally, as highlighted earlier, there is a critical need for high-quality training data accompanied by sufficient system excitation.

In summary, the incorporation of unmeasured disturbance dynamics are important to improve the quality of conventional system identification for real buildings. Additionally, an additional model for unmeasured disturbances is essential for accurately predicting future outcomes. While previous studies have demonstrated promising prediction results with their data, further investigation is needed to comprehend the variances among different system identifications and the effects of gray-box model quality and unmeasured disturbance structures on predictions. This necessity arises from the fact that gray-box model quality cannot be assured in real buildings due to sensor scarcity and low data quality.

In this research, we introduce a hybrid approach that combines a neural network-based machine-learning model for unmeasured disturbances with a gray-box model for predictive applications. Firstly, we conduct a simulation case study to underscore the deficiencies of conventional identification algorithms in handling unmeasured disturbances. We then propose and compare alternative identification algorithms to mitigate these challenges. Subsequently, we examine the limitations of these alternatives in predictive applications and propose a methodology for designing and selecting the unmeasured disturbance model. This methodology aims to prevent overfitting by considering the input-output structure of the gray-box model. Finally, we assess the performance of our proposed model on both generated simulation and experimental data.

\section{Comparisons of system identification approaches under unmeasured disturbance and effect on prediction: simulation case study}
\label{sec:body2}

\begin{figure} 
    \centering 
    \includegraphics[width=\textwidth]{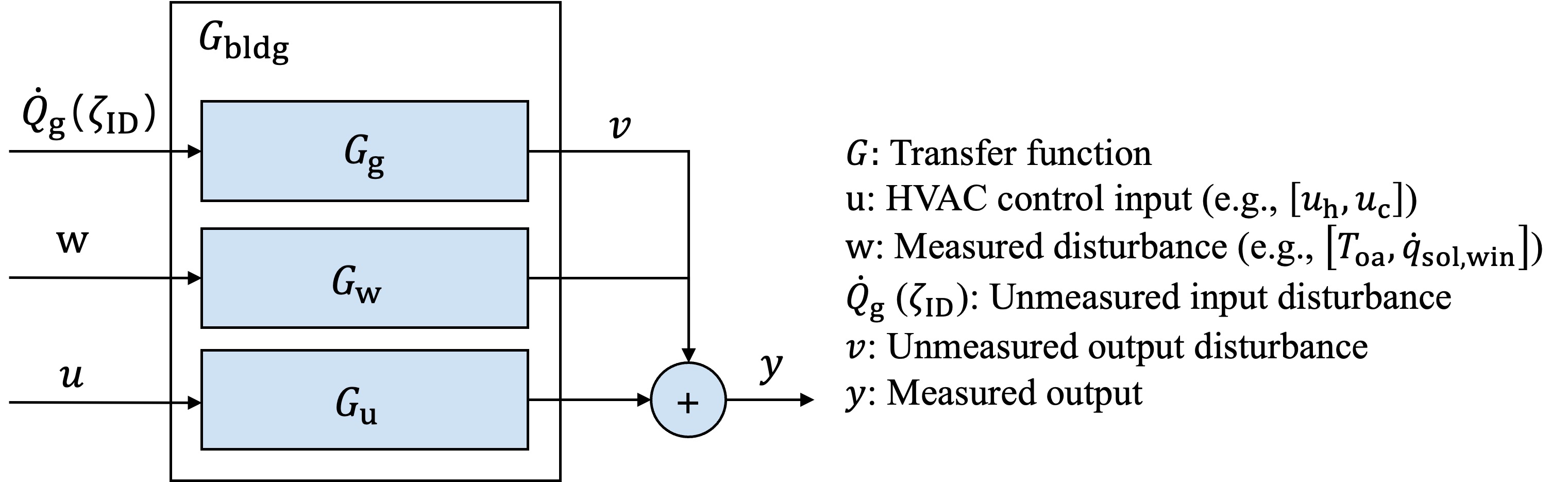} 
    \caption{Relationship between inputs and output in a building transfer function.} 
    \label{fig:tf} 
\end{figure}

In this section, a case study is presented to compare the performance of various system identification approaches when significant unmeasured disturbances are present. Fig.~\ref{fig:tf} shows the relationship between the inputs and output of a building envelope system transfer function, $G_\text{bldg}:\,\left(\textbf{w},\,\textbf{u},\, \dot{Q}_\text{g}\right)\rightarrow y$. As described in section~\ref{sec:introduction}, $\dot{Q}_\text{g}$ is not available in real buildings. Therefore, the system identification methods are classified based on how they treat this unmeasured disturbance. The conventional system identification (CONV) approach treats it as white noise \cite{Braun2002-iv,Li2015-xm}. The input disturbance approach (ID) models the unmeasured disturbance as a lumped input disturbance term ($\zeta_\text{ID}$) and includes it in the system identification \cite{Coffman2018-nq}. In contrast, the output disturbance approach (OD) represents the lumped disturbance (LD) term in the system identification through the measurement process (i.e., $y$) \cite{Kim2016-sysid, Kim2018-sysid}. In the following sections, the mathematical details of each system identification approach are described, and their performances are compared.

\subsection{True system description and data generation}
\label{sec:body21}

\begin{figure} 
    \centering 
    \includegraphics[width=250pt]{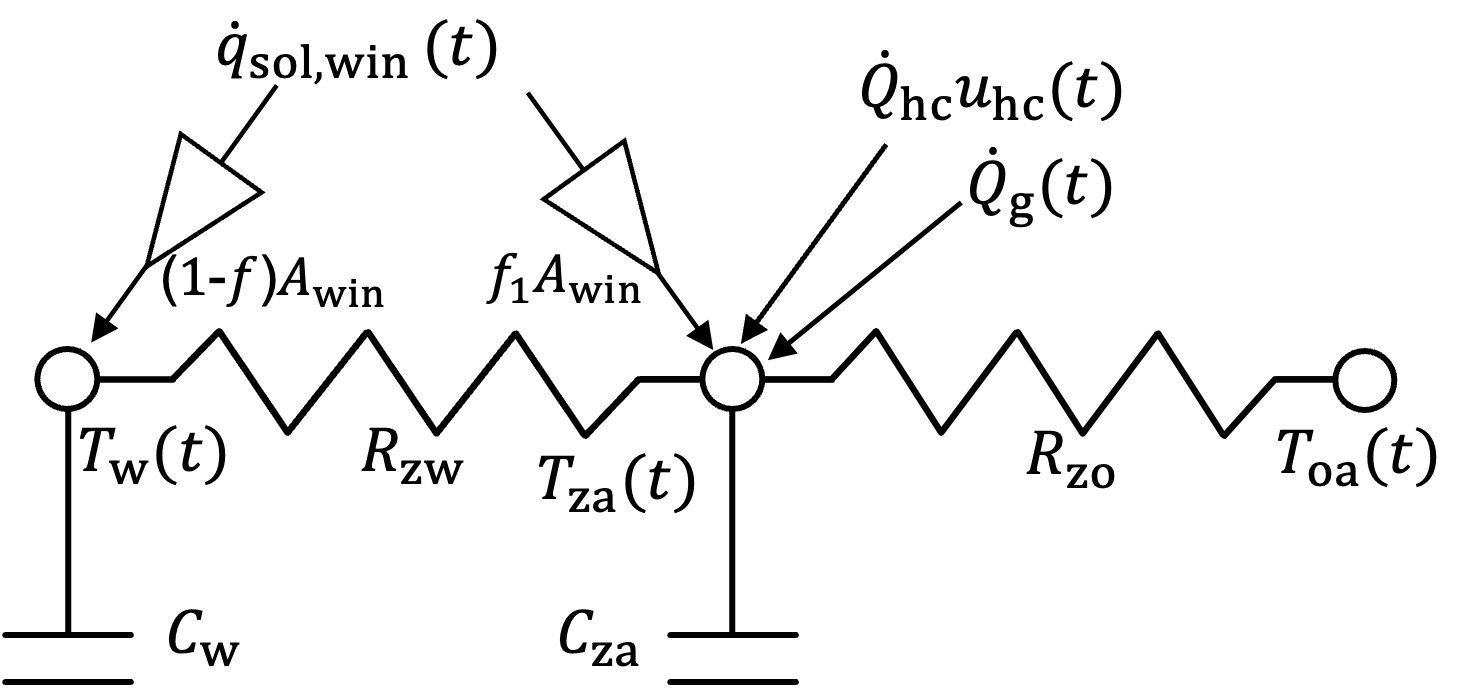} 
    \caption{A simple RC network model for a case study building.} 
    \label{fig:rc} 
\end{figure}

Evaluating the effectiveness of different system identification methods for an actual building is challenging because the true system dynamics and parameters are unknown. To address this, we developed a theoretical building envelope model, referred to as the \emph{True model} (TRUE), to generate synthetic building operational data for our analysis. The TRUE model was calibrated using a dataset from a single-zone laboratory building (FLEXLAB, \cite{Lawrence_Berkeley_National_Laboratory2021-ys}) that represents an office environment. This dataset includes weather data, thermostat setpoints, HVAC operation data, and all unmeasured disturbances such as plug loads, lighting loads, occupancy, ventilation, and infiltration. The TRUE model adopts the 2R–2C network shown in Fig.~\ref{fig:rc} and follows the state-space form given in Eq.~\ref{eq:gray_state} (state transition process) and Eq.~\ref{eq:gray_mes} (measurement process).

\begin{eqnarray}
\label{eq:gray_state}
\begin{aligned}
\underbrace{\begin{bmatrix}
\dot{T}_{\text{w}}(t)\\
\dot{T}_{\text{za}}(t)
\end{bmatrix}}_{\dot{\textbf{x}}}=&
\underbrace{\begin{bmatrix}
\frac{-1}{ C_{\text{w}} R_{\text{zw}} } & \frac{1}{ C_{\text{w}} R_{\text{zw}} } \\
\frac{-1}{ C_{\text{za}} R_{\text{zw}} } & \frac{-1}{ C_{\text{za}} R_{\text{zw}} }+\frac{-1}{ C_{\text{za}} R_{\text{zo}} } 
\end{bmatrix}}_{\textbf{A}}
\underbrace{\begin{bmatrix}
T_{\text{w}}(t)\\
T_{\text{za}}(t)
\end{bmatrix}}_{\textbf{x}}\\
&+\underbrace{\begin{bmatrix}
0 & \frac{(1-f)A_{\text{win}}}{C_{\text{w}}} \\
\frac{1}{C_{\text{za}}R_{\text{zo}}} & \frac{f A_{\text{win}}}{C_{\text{za}}} 
\end{bmatrix}}_{\textbf{B}_{\text{w}}}
\underbrace{\begin{bmatrix}
T_{\text{oa}}(t) \\
\dot{q}_{\text{sol,win}}(t)
\end{bmatrix}}_{\textbf{w}}+\underbrace{\begin{bmatrix}
0\\
 \frac{\dot{Q}_{\text{hc}}}{C_{\text{za}}}
\end{bmatrix}}_{\textbf{B}_\text{u}}
\underbrace{\begin{bmatrix}
u_{\text{hc}}(t)
\end{bmatrix}}_{\textbf{u}}+\underbrace{\begin{bmatrix}
0\\
\frac{1}{C_{\text{za}}}
\end{bmatrix}}_{\textbf{B}_{\text{g}}}
\underbrace{\begin{bmatrix}
\dot{Q}_{\text{g}}(t)
\end{bmatrix}}_{\dot{\textbf{Q}}_\text{g}}
\end{aligned}
\end{eqnarray}

\begin{eqnarray}
\label{eq:gray_mes}
y_{\text{za}}=\underbrace{\begin{bmatrix}
0&1
\end{bmatrix}}_{\textbf{C}}
\textbf{x}
\end{eqnarray}

%Cw=48/12#*3600 # 30 (kW-5min)/K 30/12*3600 kJ/K
%Cz=12/12#*3600 # 8 (kW-5min)/K
%Rzw=1.2  #K/KW  
%Rzo=9 #K/KW
%f=0.3 #convective fraction 
%Awin=3.0 # effective window area m2
%etaheat=6 # kW
%etacool=-6 # kW

The parameters to be estimated through system identification are  
$\theta=\left[ C_\text{w}, C_\text{za}, R_\text{zw}, R_\text{zo}, f, A_\text{win},\dot{Q}_\text{h},\dot{Q}_\text{c} \right]$
and were tuned using the complete set of measurements, including $\dot{Q}_\text{g}$. The TRUE model parameter values were set as follows: $C_\text{w}$ = 4.0 $\text{kWh/K}$, $C_\text{za}$ = 1.0 $\text{kWh/K}$, $R_\text{zw}$ = 1.2 $\text{K/kW}$, $R_\text{zo}$ = 9 $\text{K/kW}$, $f$ = 0.3, $A_\text{win}$ = 3.0 $\text{m}^2$, $\dot{Q}_\text{h}$ = 6 kW, and $\dot{Q}_\text{c}$ = –6 kW. An ON/OFF controller was implemented to maintain the indoor temperature, with a minimum ON/OFF duration of 5 minutes.

\begin{figure} 
    \centering 
    \includegraphics[width=\textwidth]{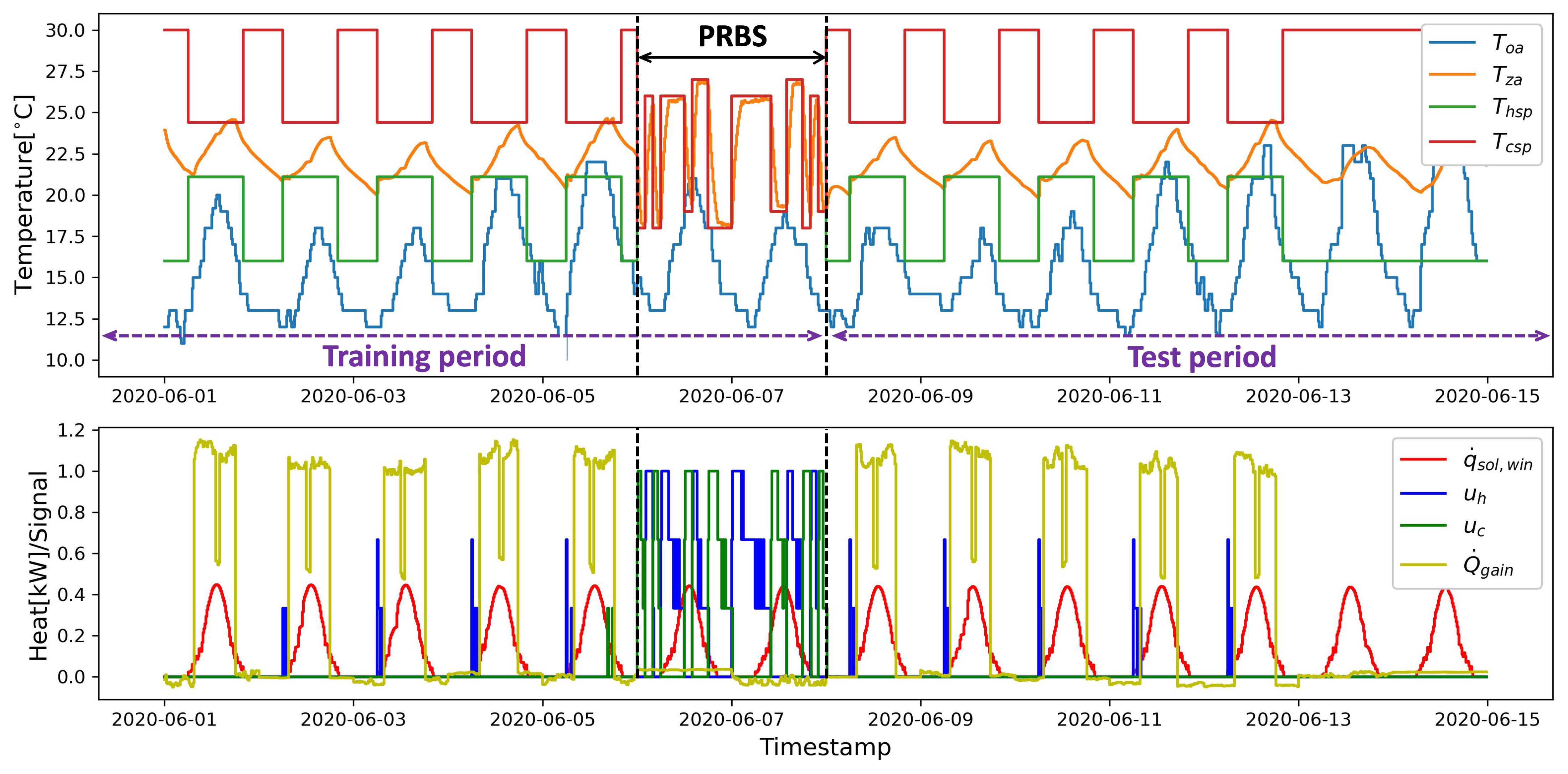} 
    \caption{Synthetic data from the TRUE model to evaluate the performance of different identification algorithms (15-minute moving average).} 
    \label{fig:sysid_data} 
\end{figure}

A two-week dataset was generated from the TRUE model using Oakland, CA weather data \cite{US_Department_of_Energy2021-fs}, as shown in Fig.~\ref{fig:sysid_data}. Here, $T_{\text{oa}}$ is the outdoor air temperature, $T_\text{za}$ is the zone air temperature, $T_\text{csp}$ is the room cooling setpoint, $T_\text{hsp}$ is the room heating setpoint, $\dot{q}_{\text{sol,win}}$ is the incident solar radiation per unit area on a zone window, $u_\text{h}$ and $u_\text{c}$ are the heating and cooling ON/OFF signals (fraction values in Fig.\ref{fig:sysid_data} indicate 15-minute moving averages), and $\dot{Q}_\text{gain}$ is the sum of all unmeasured disturbances (i.e., plug loads, lighting loads, occupancy, ventilation, and infiltration).

During weekdays, the cooling setpoint is set to $23^\circ\text{C}$–$25^\circ\text{C}$ during occupied hours (6:00–19:00) and $28^\circ\text{C}$–$30^\circ\text{C}$ during unoccupied hours. In the data generation process, internal heat gains from plug loads, lighting, occupancy, ventilation, and infiltration are included as a lumped term ($\dot{Q}_\text{g}$) with stochastic random variations. However, to test the performance of the CONV, ID, and OD system identification algorithms under a realistic scenario, $\dot{Q}_{\text{g}}$ is assumed to be unknown.  

It is further assumed that experiments can be designed to actively control the indoor temperature setpoint (within an acceptable room air temperature range) during weekends (i.e., unoccupied periods) to improve system identification. This is a practical scenario when conducting system identification in real buildings \cite{Ham2023-rg}. On the first weekend (two days), the setpoint was perturbed according to a pseudo-random binary sequence (PRBS) with a 2‑hour time scale and 4th order \cite{Madsen1993-wf}. The binary signal was mapped to sampled setpoints between $18^\circ\text{C}$ and $25^\circ\text{C}$.  

\subsection{Descriptions of  system identification approaches}
\label{sec:body22}

Eqs.~\ref{eq:gray_state}–\ref{eq:gray_mes} were discretized with a 15‑minute sampling interval using a zero‑order hold method \cite{Rouchier2019-uc}. Three different system identification approaches (CONV, ID, and OD) were then applied to the generated data. While all three approaches share the same discretized model structure for the building envelope dynamics, they differ in the structure of their disturbance models. It is important to note that the unmeasured disturbance ($\dot{Q}_{\text{g}}$) was not provided to the identification algorithms. In other words, the identified model $G$ maps only the measured disturbances ($T_\text{oa}$, $\dot{q}_{\text{sol,win}}$) and the control input ($\dot{Q}_{\text{hc}}$) to the indoor air temperature ($y_\text{za}$), as shown in Fig.~\ref{fig:tf}.  

\subsubsection{Conventional simulation error minimization approach}

The CONV approach assumes that all the unmeasured disturbance can be expressed as white noise ($e_{\text{conv}}: {\varepsilon}_{\text{conv}}(k)\sim N(0,\sigma^2_{\text{conv}})$) in the measurement process, and the discretized system can be written as Eq. (\ref{eq:conv_con});
\begin{eqnarray}
\begin{aligned}
\label{eq:conv_con}
\textbf{x}(k+1)&=\textbf{A}_{\text{d}}\textbf{x}(k)+\textbf{B}_{\text{w,d}}\textbf{w}(k)+\textbf{B}_{\text{u,d}}\textbf{u}(k)\\
y(k)&=\textbf{C}_{\text{d}}\textbf{x}(k)+e_{\text{conv}}(k)
\end{aligned}
\end{eqnarray}
The set of parameters ($\theta_{\text{conv}}^{*}$) is estimated by minimizing the sum of squared errors between simulation ($\hat{y}(k;\theta)=\textbf{C}_{\text{d}}\hat{\textbf{x}}(k;\theta)$) and measurement ($y(k)$) via the nonlinear optimization (Eqs. \ref{eq:conv_dis}-\ref{eq:conv_sysid});
\begin{eqnarray}
\begin{aligned}
\label{eq:conv_dis}
\hat{\textbf{x}}(k+1;\theta)&=\textbf{A}_{\text{d}}(\theta)\hat{\textbf{x}}(k;\theta)+\textbf{B}_{\text{w,d}}(\theta)\textbf{w}(k)+\textbf{B}_{\text{u,d}}(\theta)\textbf{u}(k)\\
y(k)&=\textbf{C}_{\text{d}}\hat{\textbf{x}}(k;\theta)+{\varepsilon}_{\text{conv}}(k;\theta)
\end{aligned}
\end{eqnarray}
\begin{eqnarray}
\label{eq:conv_sysid}    
\theta_{\text{conv}}^{*}=\arg\min_{\theta}\sum_{k=1}^{N}\left({\varepsilon}_{\text{conv}}(k;\theta)\right)^2,  \end{eqnarray}
where subscript $\text{d}$ indicates a discretized system and, $k$ is a discrete time step.

In the 7-day training data, the initial state (i.e., $\textbf{x}(0)$) is obtained via a Kalman filter \cite{Rouchier2019-uc} by using the first-day data. Then, the following 6 days are predicted via simulation (Eq. \ref{eq:conv_dis}). The optimization bounds of parameters are set to $\left[0.1,\,40 \right]$ for all $R$ and $C$ parameters (i.e., $C_\text{w},C_\text{za},R_\text{zw},R_\text{zo}$), $\left[\text{1e-6},\, 1\right]$ for $f$, and $\left[0.1,\, 25\right]$ for $A_\text{win}$.

\subsubsection{Input disturbance identification approach}

The ID approach assumes that unmeasured disturbances come from the input channel, i.e., the heat gain term, and treats the input disturbance as an additional dynamic state. This can be written as an augmented state space format (Eq. \ref{eq:ID_con}) \cite{Coffman2018-nq,Radecki2017-wj}.
\begin{eqnarray}\label{eq:ID_con}
\begin{aligned}
\underbrace{\begin{bmatrix}
\dot{\textbf{x}}\\
\dot{\zeta}_\text{ID}
\end{bmatrix}}_{\dot{\textbf{x}}_{\text{ID}}}&=
\underbrace{\begin{bmatrix}
\textbf{A}& \textbf{A}_{\zeta_{\text{ID}}} \\
\textbf{0}&\textbf{0}
\end{bmatrix}}_{\textbf{A}_{\text{ID}}}
\underbrace{\begin{bmatrix}
\textbf{x}\\\zeta_\text{ID}
\end{bmatrix}}_{\textbf{x}_\text{ID}}+\textbf{B}_\text{w}\textbf{w}+\textbf{B}_\text{u}\textbf{u}+\textbf{e}_{\text{x}_\text{ID}},\,\,\text{and }\textbf{A}_{\zeta_{\text{ID}}}=
\begin{bmatrix}
\textbf{0}\\\frac{1}{C_{\text{za}}}
\end{bmatrix}\\
y_{\text{za}}&=\underbrace{\begin{bmatrix}
\textbf{C}&0
\end{bmatrix}}_{\textbf{C}_{\text{ID}}}\textbf{x}_{\text{ID}}+e_{\text{ID}}
\end{aligned}
\end{eqnarray}
where $\textbf{e}_{\textbf{x}_\text{ID}}$ and $e_{\text{ID}}$ are state and measurement noises. $\zeta_{\text{ID}}$ represents the lumped input disturbance term.

The key idea in this approach is to treat $\zeta_{\text{ID}}$ as Wiener process, so therefore, it behaves as the Brownian motion after a discretization according to its noise level (i.e., $\zeta_\text{ID}(k+1)=\zeta_\text{ID}(k)+\varepsilon_{\zeta_\text{ID}}(k|\theta)$).

For the system identification, the ID approach first calculates one-step-ahead prediction errors via the following steps. The system equation (Eq.~\ref{eq:ID_con}) is discretized as Eq.~\ref{eq:ID_dis}. From the initial states ($\hat{\textbf{x}}_{\text{ID}}(1|1)$), the next time states ($\hat{\textbf{x}}_{\text{ID}}(2|1)$) and zone air temperature ($\hat{y}_{\text{za}}(2|1)$) are predicted through Eq. \ref{eq:ID_con}. The prediction error (i.e., innovation, $\varepsilon_{\text{ID}}$) is estimated through Eq.~\ref{eq:ID_inno}, and then, the predicted states are updated by using the innovation and optimal Kalman gain ($\textbf{K}(k|\theta)$) (Eq. \ref{eq:ID_filter}). The process in Eqs. \ref{eq:ID_dis}-\ref{eq:ID_filter} is sequentially repeated for the whole data ($k=1,2,...,N$). The ID approach finds a set of parameters by minimizing the square sum of one-step ahead prediction error (Eq. \ref{eq:ID_sysid}).
\begin{eqnarray}
\label{eq:ID_dis}
\begin{aligned}
\hat{\textbf{x}}_{\text{ID}}(k+1|k;\theta)&=\textbf{A}_{\text{d},\text{ID}}(\theta)\hat{\textbf{x}}_{\text{ID}}(k|k)+\textbf{B}_{\text{w,d}}(\theta)\textbf{w}(k)+\textbf{B}_{\text{u,d}}(\theta)\textbf{u}(k)\\
\hat{y}_{\text{ID}}(k+1|k;\theta)&=\textbf{C}_{\text{d},\text{ID}}\hat{\textbf{x}}_{\text{ID}}(k+1|k)
\end{aligned}
\end{eqnarray}

At each time step $k$, the optimal Kalman gain is obtained using the Kalman filter \cite{Rouchier2019-uc}. The gain is estimated based on the state noise covariance ($\boldsymbol{\Sigma}_{\textbf{x}_\text{ID}}$, i.e., $\boldsymbol{\varepsilon}_{\text{x}_\text{ID}}(k) \sim N(0,\boldsymbol{\Sigma}_{\textbf{x}_\text{ID}})$) and the measurement noise covariance $\boldsymbol{\Sigma}_{y_\text{za}}$ in the discretized system. The measurement noise covariance is set to $0.25^2/T_\text{s}$, based on the sensor noise level ($\pm 0.5 ^\circ \text{C}$) and the discretization sampling time $T_\text{s}$. The state noise covariance is modeled using two additional parameters, i.e., $\boldsymbol{\Sigma}_{\textbf{x}_\text{ID}} = \text{diag}(\sigma_x^2, \sigma_x^2, \sigma_{\zeta_{\text{ID}}}^2)$, which determine the degree of state update in $\hat{\textbf{x}}_{\text{ID}} = \left[\hat{\textbf{x}}, \hat{\zeta}_{\text{ID}}\right]^\intercal$ (Eq.~\ref{eq:ID_filter}). The optimization bounds of these two parameters are set to [1e$^{-9}$, 1].  
\begin{eqnarray}
\label{eq:ID_inno}
\varepsilon_{\text{ID}}(k+1;\theta)=y(k+1)-\hat{y}_{\text{ID}}(k+1|k;\theta)
\end{eqnarray}
\begin{eqnarray}
\label{eq:ID_filter}
\hat{\textbf{x}}_{\text{ID}}(k+1|k+1;\theta)=\hat{\textbf{x}}_{\text{ID}}(k+1|k;\theta)+\textbf{K}(k+1;\theta)\varepsilon_{\text{ID}}(k+1;\theta)
\end{eqnarray}
\begin{eqnarray}
\label{eq:ID_sysid}    
\theta_{\text{ID}}^{*}=\arg\min_{\theta}\sum_{k=1}^{N}\left(\varepsilon_{\text{ID}}(k;\theta) \right)^2.     
\end{eqnarray}

\subsubsection{Output disturbance identification approach}
\label{sec:body223}
The OD approach \cite{Kim2016-sysid,Kim2018-sysid} does not explicitly model the input disturbances. Instead, it tries to model the effect of unmeasured heat gains on the output (i.e., room air temperature). The aggregated contribution of the unknown heat sources to the output is called the output disturbance, as opposed to the input disturbance. The OD approach models the output disturbance as a filtered process of white noise ($e_{\text{OD}}(k)$), which is called output disturbance ($v_{\text{OD}}(k)$ in Eq. \ref{eq:od_dis}) and Fig. \ref{fig:tf}. The output disturbance dynamics can be modeled with two more parameters, $\rho_1$ and $\rho_2$ (Eq. \ref{eq:od_dis}).

\begin{eqnarray}
\label{eq:od_dis}
\begin{aligned}
\textbf{x}(k+1)&=\textbf{A}_\text{d}\textbf{x}(k)+\textbf{B}_\text{w,d}\textbf{w}(k)+\textbf{B}_\text{u,d}\textbf{u}(k)\\
y(k)&=\textbf{C}_\text{d}\textbf{x}(k)+v_{\text{OD}}(k)\\
\zeta_\text{OD}(k+1)&=\underbrace{\begin{bmatrix}\rho_1 \end{bmatrix}}_{\mathcal{F}}\zeta_\text{OD}(k)+\underbrace{\begin{bmatrix}\rho_2 \end{bmatrix}}_{\mathcal{G}}e_{\text{OD}}(k)\\
v_{\text{OD}}(k)&=\zeta_\text{OD}(k)+e_{\text{OD}}(k)
\end{aligned}
\end{eqnarray}

For each time step, the prediction error (i.e., innovation, $\varepsilon_{\text{OD}}$) is estimated via Eq. \ref{eq:od_inno}. Then it is used to calculate next time prediction (Eq. \ref{eq:od_pred}-\ref{eq:od_obj}).

\begin{eqnarray}
\label{eq:od_inno}
\varepsilon_{\text{OD}}(k;\theta)=y(k)-\hat{y}_{\text{OD}}(k;\theta)
\end{eqnarray}

\begin{eqnarray}
\label{eq:od_pred}
\begin{aligned}
\hat{\zeta}_\text{OD}(k+1;\theta)&=\mathcal{F}(\theta)\hat{\zeta}_\text{OD}(k;\theta)+\mathcal{G}(\theta)\varepsilon_{\text{OD}}(k)\\
\hat{v}_{\text{OD}}(k;\theta)&=\hat{\zeta}_\text{OD}(k;\theta)+\varepsilon_{\text{OD}}(k;\theta)\\
\hat{\textbf{x}}(k+1;\theta)&=\textbf{A}_\text{d}(\theta)\hat{\textbf{x}}(k;\theta)+\textbf{B}_{\text{w,d}}(\theta)\textbf{w}(k)+\textbf{B}_{\text{u,d}}(\theta)\textbf{u}(k)\\
\hat{y}_{\text{OD}}(k+1;\theta)&=\textbf{C}_\text{d}\hat{\textbf{x}}(k+1;\theta)+\mathcal{F}(\theta)\hat{\zeta}_\text{OD}(k+1;\theta)
\end{aligned}
\end{eqnarray}

The optimal set of parameters is estimated by minimizing the square sum of one-step ahead prediction error (Eq. \ref{eq:od_obj}). The optimization bounds of $\rho_1\text{ and }\rho_2$ are set to [-0.999, 0.999] as suggested in \cite{Kim2016-sysid}.

\begin{eqnarray}
\label{eq:od_obj}    
\theta_{\text{OD}}^{*}=\arg\min_{\theta}\sum_{k=1}^{N}\left(\varepsilon_{\text{OD}}(k;\theta)\right)^2
\end{eqnarray}

\subsection{System identification results and discussion}
\label{sec:body23}
For each identification algorithm, the optimization problem is non-convex, and therefore, we randomly sampled initial starting points and repeatedly solved the optimization problems 50 times to find a better optimal solution. The estimation results of  CONV, ID, OD identification approaches are summarized in Table \ref{tb:tb1_theta}.

\begin{table}[!htbp]
\centering
\begin{tabularx}{\textwidth}{@{}Y|YYYYYYYYYY@{}}
\toprule
&$C_\text{w}$ [kWh/K] &$C_\text{za}$ [kWh/K] &$R_\text{zw}$ [K/kW]&$R_\text{zo}$ [K/kW]&$f$ [-]&$A_\text{win}$ [m$^2$]&$\dot{Q}_\text{h}$ [kW]&$\dot{Q}_\text{c}$ [kW]&$\sigma_x$(ID) $\rho_1$(OD)&$\sigma_{\zeta}$(ID) $\rho_2$(OD)\\
\midrule
TRUE & 4.0  & 1.0 & 1.2 & 9.0 & 0.3 &3.0&6.0&-6.0& & \\ 
CONV& 40 & 0.4 & 6.2 & 40 & 1.0  &1.1&1.5&-1.3&&  \\ 
ID& 20.6 & 1.4 & 0.8 & 3.4 & 0.03 & 20.0&6.7&-8.7 &1e-3& 2e-2  \\ 
OD&  10.6 & 1.3 & 0.8 & 4.1 & 0.2 &12.4&6.7&-8.6&0.99&0.99 \\ 
\bottomrule
\end{tabularx}
\label{tb:tb1_theta}
\caption{Comparison of estimated parameters from system identification approaches}
\end{table}

Table~\ref{tb:tb1_theta} shows that the CONV method yielded inaccurate parameter estimates. In particular, it significantly overestimated the thermal capacitance ($C_{\text{w}}$) and the thermal resistance ($R_{\text{zo}}$). This occurred because the method attempted to explain the relationship between inputs and outputs without considering heat gain information, leading to compensatory changes in the physical parameters. As a result, the large thermal mass ($C_\text{w}$) acted like a heat source during the day to offset the unmeasured heat gains, while $R_{\text{zo}}$ was overestimated to retain heat in the thermal mass overnight. In addition, the estimated heating and cooling rates ($\dot{Q}_{\text{h}}$ and $\dot{Q}_{\text{c}}$) were substantially underestimated, producing values that do not match the physical scales of the actual heating and cooling capacities.  

In contrast, the ID and OD algorithms include an additional degree of freedom through parameters associated with the disturbance models, allowing them to better explain the input–output relationship under unmeasured disturbances. Overall, incorporating disturbance dynamics improves parameter estimation when such disturbances are present.  

The ID method produced improved estimates of $R_{\text{zo}}$, indicating its ability to capture small unmeasured disturbances. However, it struggled with abrupt changes in these disturbances, such as during transitions between occupied and unoccupied periods. In such cases, the method compensated by overestimating the solar radiation parameters ($A_{\text{win}}$ and $f$). The OD method, on the other hand, provided more accurate estimates for most parameters, although they were not entirely precise. As noted in \cite{Kim2016-sysid}, input disturbances are often noisier and more volatile than output disturbances, resulting in a higher-frequency power spectrum. In contrast, output disturbances tend to have a higher power spectrum in the lower-frequency range, which can be easier to model in certain applications.

\begin{figure} 
    \centering 
    \includegraphics[width=\textwidth]{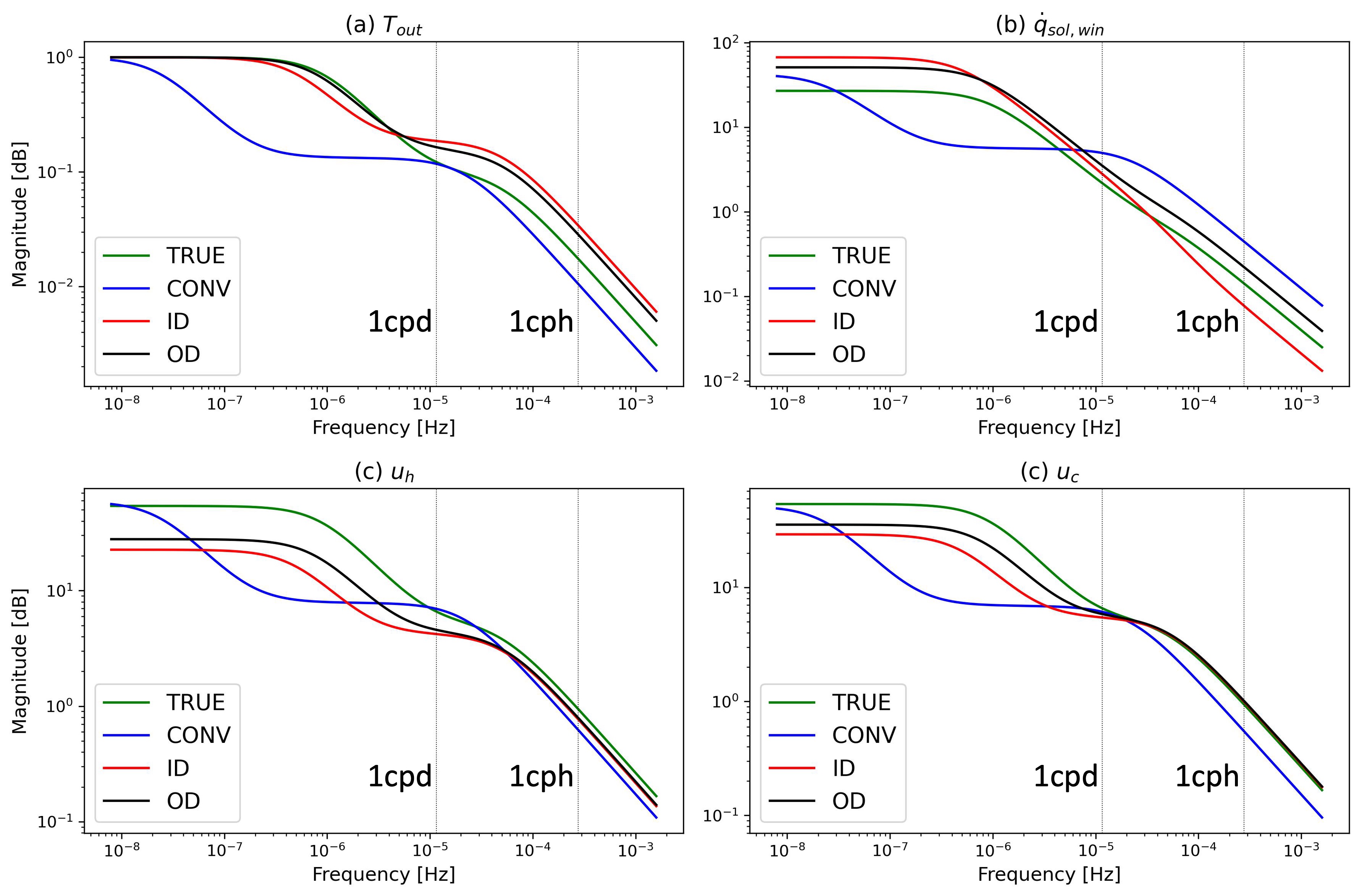} 
    \caption{Comparison of Bode magnitude plots of measured disturbances ($T_{\text{oa}}$ and $\dot{q}_{\text{sol,win}}$) and control inputs ($u_\text{h}$ and $u_\text{c}$) for each method. (1 cph = 2.8e-4 Hz, 1 cpd = 1e-5 Hz)} 
    \label{fig:bode} 
\end{figure}

Since all methods deviate from the TRUE model, it is important to evaluate their ability to capture the building’s thermal dynamics and to identify potential issues in prediction applications. To this end, Bode magnitude plots of the different approaches are compared with the TRUE model in Fig.~\ref{fig:bode}. Both the ID and OD models outperform the CONV method overall. Specifically, the CONV method failed to capture the behavior of $u_{\text{h}}$ and $u_{\text{c}}$ in the high-frequency range, whereas the ID and OD approaches performed better in this regard. This suggests that short-term heating and cooling behaviors can be captured more effectively with ID and OD, highlighting the importance of selecting an appropriate identification method for specific predictive applications.

\begin{figure} 
    \centering 
    \includegraphics[width=\textwidth]{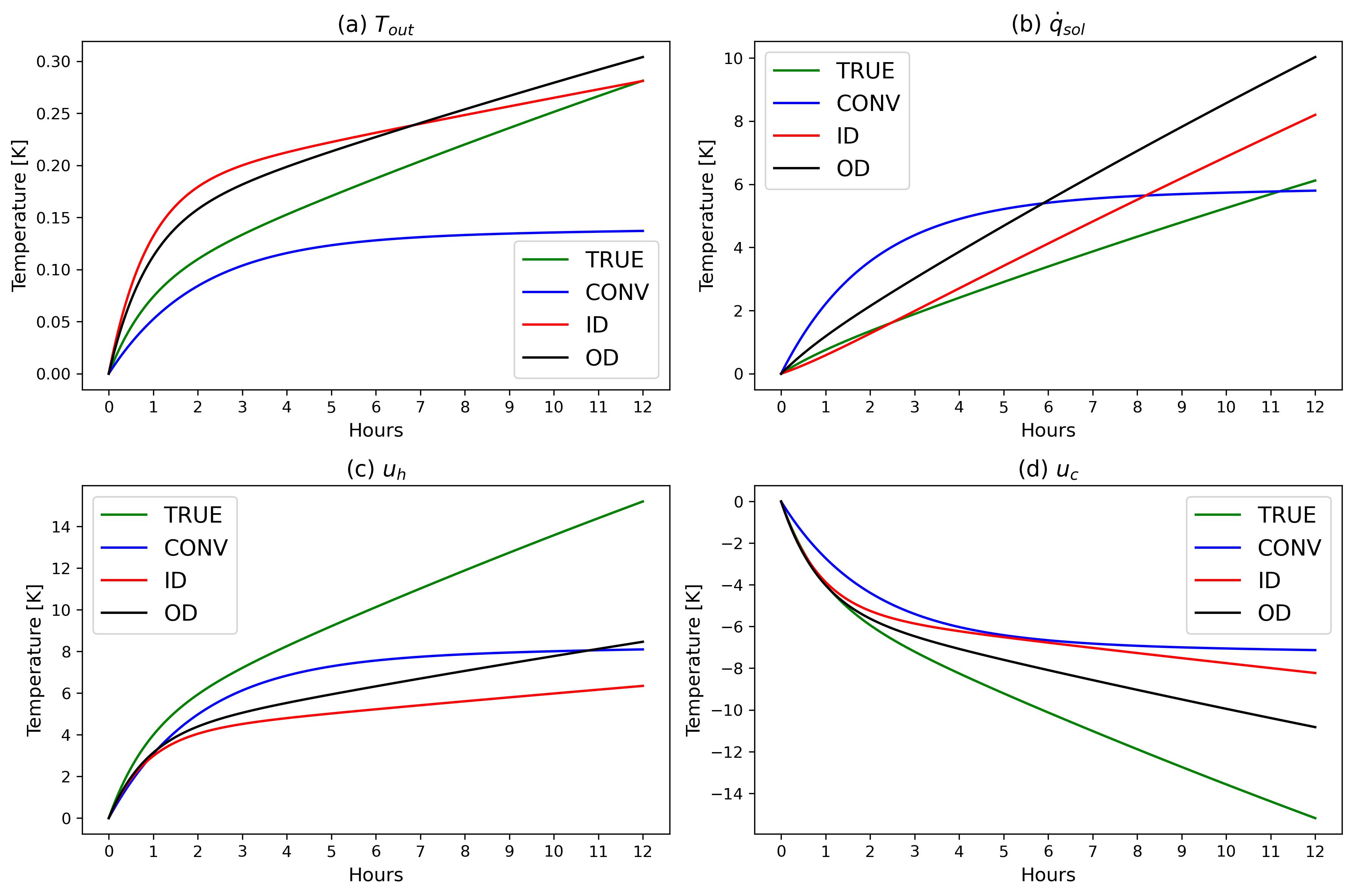} 
    \caption{Comparison of step response of measured disturbances ($T_{\text{oa}}$ and $\dot{q}_{\text{sol,win}}$) and control input ($u_\text{h}$ and $u_\text{c}$) for each method.} 
    \label{fig:step} 
\end{figure}

The step responses of the measured disturbances ($T_{\text{oa}}$ and $\dot{q}_{\text{sol,win}}$) and control inputs ($u_\text{h}$ and $u_\text{c}$) over a 12-hour period are compared with the TRUE system in Fig. \ref{fig:step}. For the heating control input, all methods perform well up to 30 minutes but show a rapid decline thereafter. For the cooling control input, ID and OD maintain good performance during the first 1.5 hours, whereas CONV performs poorly even over a short-term horizon.  

\begin{figure} 
    \centering 
    \includegraphics[width=\textwidth]{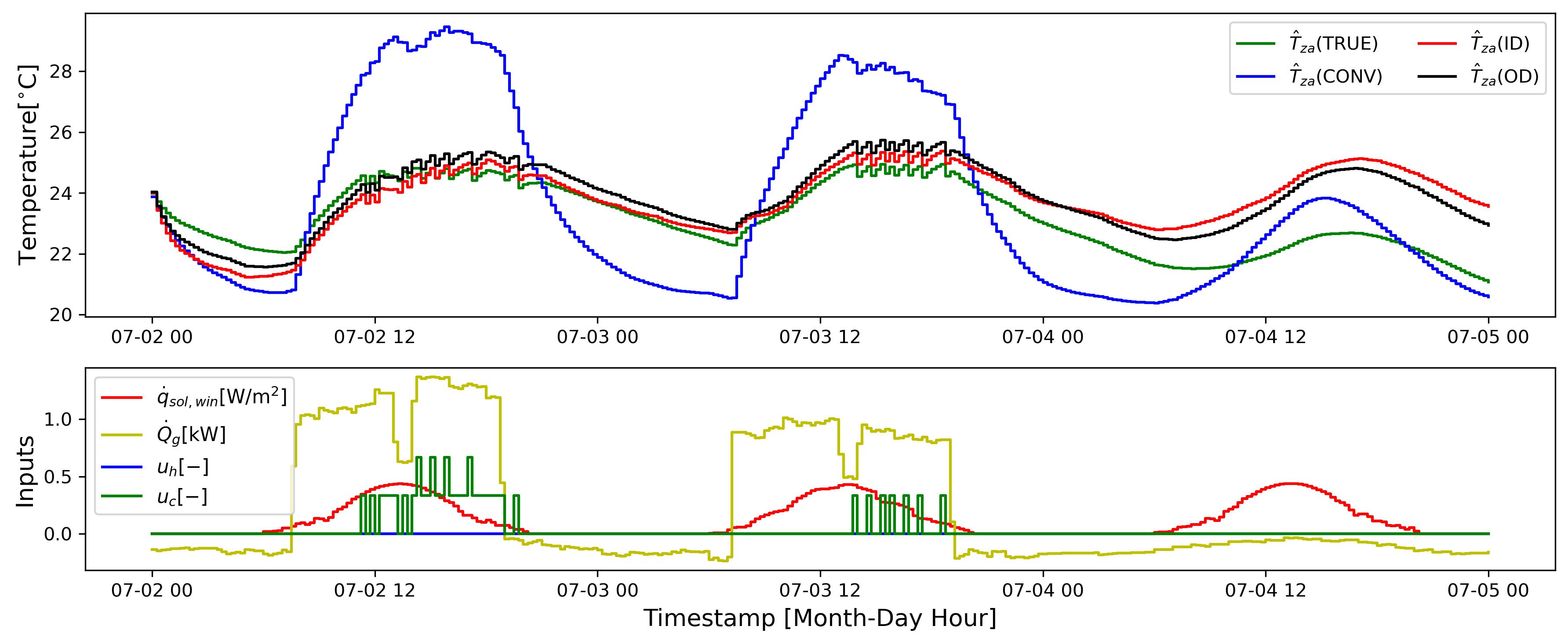} 
    \caption{Comparison of indoor temperature predictions with unmeasured disturbance as an input for each method.} 
    \label{fig:nstep} 
\end{figure}

To compare the performance of each method, three-day-ahead temperature predictions using unmeasured disturbances as input data are evaluated (Fig.~\ref{fig:nstep}). As expected, the ID and OD methods show superior performance compared to the CONV method. Since the CONV method embeds all unmeasured disturbance information in its parameters, its temperature predictions deviate significantly when the magnitude of the unmeasured disturbance is large. Although small deviations are observed, the predictions of the ID and OD methods are generally close to the TRUE predictions. However, their performance declines during the weekend, when large unmeasured disturbances are absent, as their parameters perform poorly in capturing slow and long-term dynamics (i.e., the low-frequency region in Fig.~\ref{fig:bode}).

For evaluating model accuracy, we compared the predicted heating and cooling loads with measurements, in addition to using the typical cross-validation strategy that directly compares output predictions (in our case, thermostat temperatures) with measurements from a validation dataset. This is important for control applications (e.g., MPC), where accurately estimating the required heating or cooling rate is essential—more specifically, the runtime fraction (RTF) of the heating and cooling stages in our case, i.e., $\bar{u}_\text{h}(k)$ and $\bar{u}_\text{c}(k)$. To achieve this, the state-space model in Eqs. \ref{eq:gray_state}–\ref{eq:gray_mes} was rewritten as Eqs. \ref{eq:inverseQ1}–\ref{eq:inverseQ2}, ignoring the error noise term. The required RTF ($\hat{\textbf{u}}(k)$) and the next-time states ($\hat{\textbf{x}}(k+1)$) were estimated using Eq. \ref{eq:inverseQ2}. In this case, $\left(\textbf{C}_{\text{d}}\textbf{B}_{\text{d,u}}(\theta)\right)$ is not invertible, so its pseudo-inverse, $\left(\textbf{C}_\text{d}\textbf{B}_{\text{d,u}}(\theta)\right)^{\dagger}$, was used.  

\begin{eqnarray}
\begin{aligned}
\label{eq:inverseQ1}
\textbf{y}(k+1)&=\textbf{C}_\text{d}(\theta)\textbf{x}(k+1)\\
&=\textbf{C}_\text{d}(\theta)\left( \textbf{A}_\text{d}\textbf{x}(k)+\textbf{B}_{\text{d,u}}(\theta)\textbf{u}(k)+\textbf{B}_{\text{d,w}}(\theta)\textbf{w}(k) \right)
\end{aligned}
\end{eqnarray}

\begin{eqnarray}
\begin{aligned}
\label{eq:inverseQ2}
\hat{\textbf{u}}(k)&=\left(\textbf{C}_\text{d}\textbf{B}_{\text{d,u}}(\theta)\right)^{\dagger}\left[\textbf{y}(k+1)-\textbf{C}_\text{d}(\theta)\left(\textbf{A}_\text{d}\hat{\textbf{x}}(k)+\textbf{B}_{\text{d,w}}(\theta)\textbf{w}(k)\right)\right]\\
\hat{\textbf{x}}(k+1)&=\textbf{A}_\text{d}\hat{\textbf{x}}(k)+\textbf{B}_{\text{d,u}}(\theta)\hat{\textbf{u}}(k)+\textbf{B}_{\text{d,w}}(\theta)\textbf{w}(k) \\
\end{aligned}
\end{eqnarray}

%\begin{eqnarray}
%\label{eq:14}
%\begin{aligned}
%\begin{bmatrix}
%\dot{T}_\text{w}
%\end{bmatrix}&=\begin{bmatrix}
%\frac{-1}{C_\text{w} R_{\text{zw}}}
%\end{bmatrix}\begin{bmatrix}
%T_\text{w}
%\end{bmatrix}+\begin{bmatrix}
%0&\frac{(1-f)A_\text{win}}{C_{\text{w}}}&0&\frac{1}{C_\text{w} %R_{\text{zw}}}
%\end{bmatrix}\begin{bmatrix}
%T_{\text{oa}}\\\dot{q}_{\text{sol,win}}\\ %\dot{Q}_{\text{hc}}\\T_{\text{za}}
%%\end{bmatrix}\\
%\dot{Q}_{\text{hc}}(k)&=\frac{T_{\text{za}}(k)-T_{\text{w}}(k)}%{R_{\text{zw}}}\\&+\frac{T_{\text{za}}(k)-T_{\text{oa}}(k)}{R_{\text{zo}}}-fA_{\text{win}}\dot{q}_{\text{sol,win}}(k)-\dot{Q}_{\text{gain}}(k)
%\end{aligned}
%\end{eqnarray}

\begin{figure} 
    \centering 
    \includegraphics[width=\textwidth]{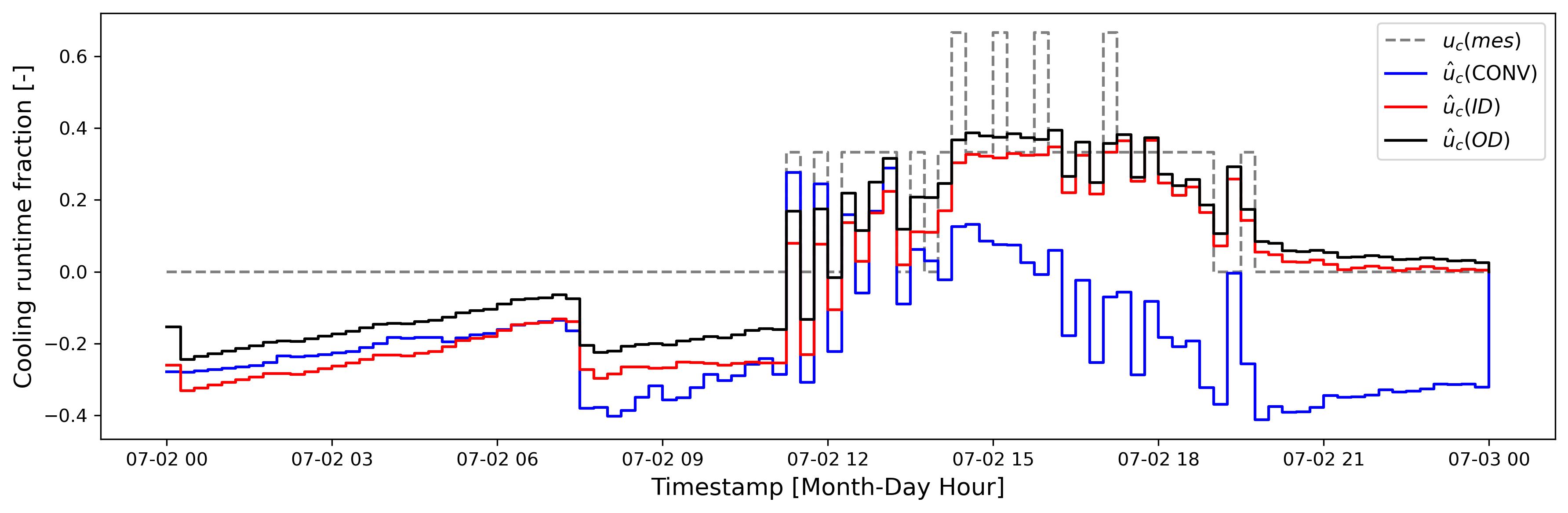} 
    \caption{Comparison of required cooling runtime fraction to maintain the measured temperature.} 
    \label{fig:invq} 
\end{figure}

In Fig.~\ref{fig:invq}, the required cooling RTF at each sampling time to maintain the measured temperature is compared across the methods. Similar to the temperature prediction results, the ID and OD methods perform well during cooling periods. However, all methods show some deviations during unoccupied periods.

To summarize, the OD method provides the best performance, although all approaches fail to capture long-term predictions. The ID method shows performance similar to the OD method, whereas the CONV method performs substantially worse than the other two, as expected. As described in \cite{Kim2016-sysid}, the output disturbance is a low-pass filtered version of the input disturbance, meaning that the smoother signal can be well modeled by the disturbance dynamics (Eq.~\ref{eq:od_dis}). Similarly, the high-frequency nature of the input disturbance can be influenced by the estimation of the noise parameter scales ($\text{e}_{\text{x}_\text{ID}}$ in Eq.~\ref{eq:ID_con}). Therefore, the OD method is adopted in this research due to both its performance and robustness.

\subsection{Challenges in prediction application}
\label{sec:body24}

In predictive applications such as MPC, accurate prediction performance is critically important. However, even with a perfect gray-box model, prediction becomes difficult when unmeasured disturbances are significant. To illustrate this, consider a building subject to large unmeasured disturbances. In such cases, predicting the building dynamics is extremely challenging because the magnitude of the unmeasured disturbance is unknown. This section presents a detailed simulation study to highlight this issue and emphasize the need for an additional modeling approach—namely, the hybrid modeling method described in the following section—to address this limitation.  

\begin{figure} 
    \centering 
    \includegraphics[width=\textwidth]{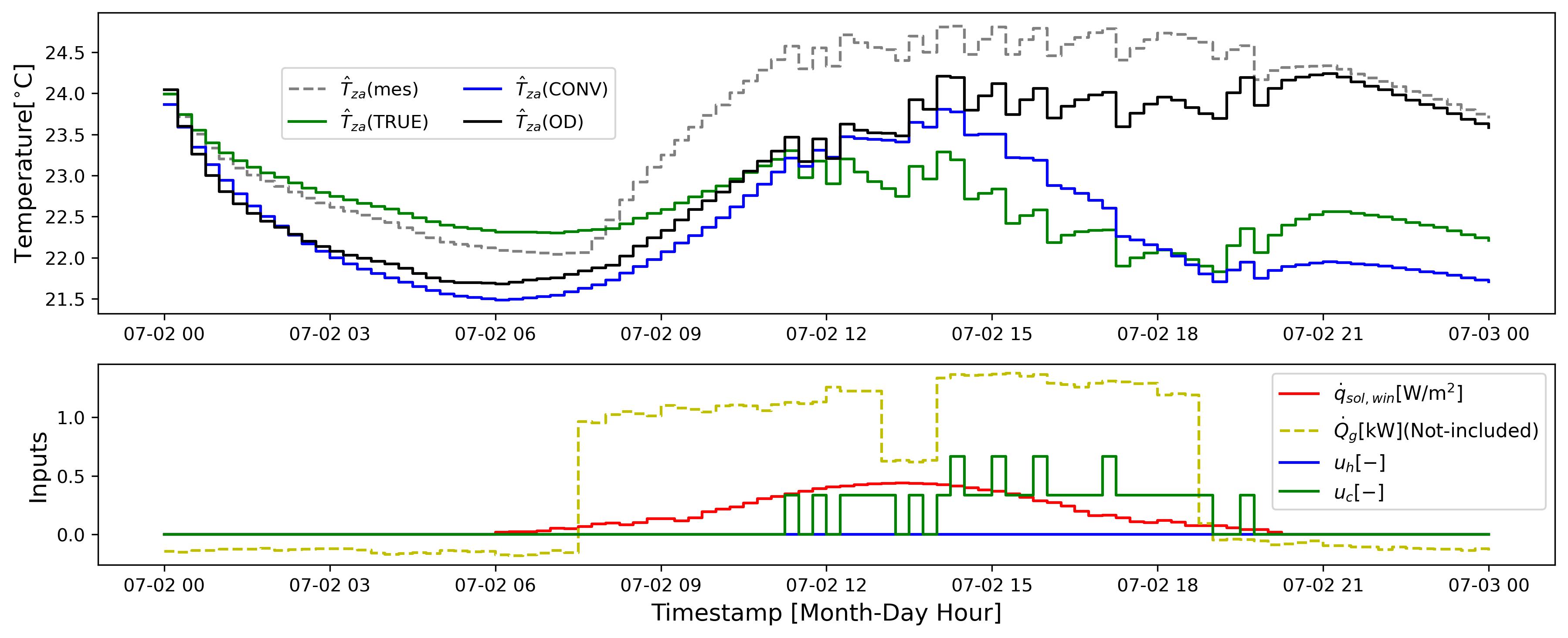} 
    \caption{1-day ahead prediction of the TRUE, CONV and OD models without heat gain information.} 
    \label{fig:nstep_woQ} 
\end{figure}

To evaluate the predictive performance of the models, we compared the one-day-ahead room air temperature predictions of the TRUE, CONV, and OD models with measurements (mes) (Fig.~\ref{fig:nstep_woQ}). Heat gain information (shown as the yellow-dotted line in the bottom figure) was deliberately excluded to simulate a realistic prediction scenario. Without this information, all models deviated from the measurements. Although OD performed better because it implicitly embeds heat gain effects in its parameters, this does not necessarily indicate superiority, as even the TRUE model produced incorrect predictions. The CONV model was expected to capture more heat gain information through its parameters, yet its predictions were not close to either the TRUE model or the measurements. These results indicate that, in addition to obtaining accurate parameters, it is essential to develop methods for forecasting unmeasured disturbances and compensating for parameter inaccuracies in predictive applications such as MPC.

\section{Hybrid modeling approach for predictive application}
\label{sec:body3}

\subsection{Overview of hybrid modeling approach}
\label{sec:body31}

\begin{figure} 
    \centering 
    \includegraphics[width=\textwidth]{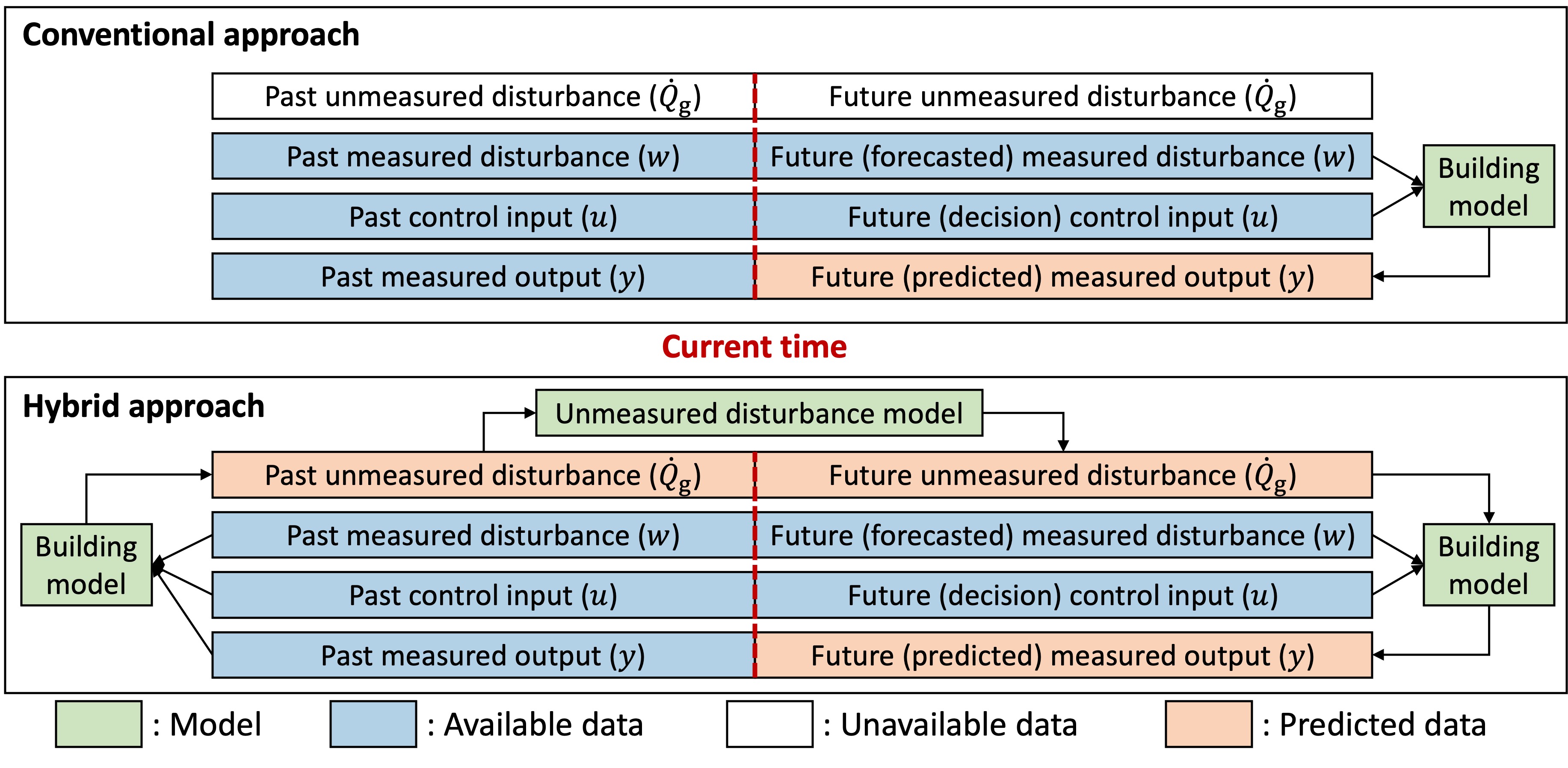} 
    \caption{Comparison between Conventional and Hybrid approach.} 
    \label{fig:hybrid_structure} 
\end{figure}

In this section, we present a hybrid modeling approach that combines a gray-box building model with a machine-learning model to forecast unmeasured disturbances for predictive applications (Hybrid approach). Figure~\ref{fig:hybrid_structure} illustrates the overall structure of the Hybrid approach in comparison with a conventional prediction approach (Conventional approach).

In the Conventional approach, the building model (i.e., a transfer function between the inputs (control inputs, measured disturbances, and unmeasured disturbances) and the output (temperature) as shown in Fig.~\ref{fig:tf}) predicts future temperature using forecasts of future measured disturbances (i.e., $T_{\text{oa}}$ and $\dot{q}_\text{sol,win}$) and future control inputs ($u_{\text{h}}$ and $u_{\text{c}}$). However, as discussed in Section~\ref{sec:body24} (see Fig.~\ref{fig:nstep_woQ}), this approach does not provide reliable predictions when the impact of unmeasured disturbances is significant, regardless of the quality of the gray-box building model.  

\begin{figure}
    \centering 
    \includegraphics[width=300pt]{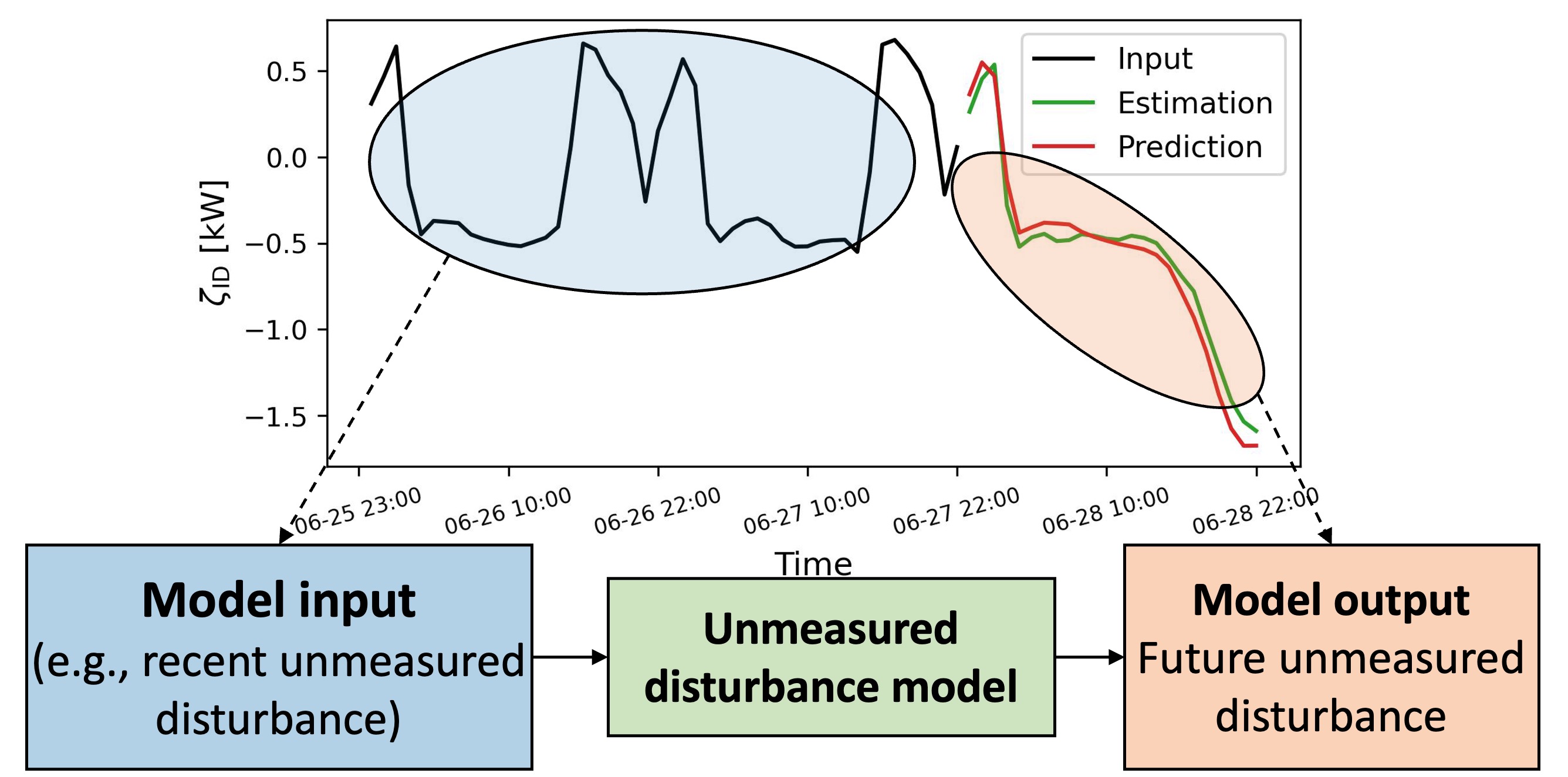} 
    \caption{A graphical example of the disturbance model (ID case).} 
    \label{fig:hybrid_example} 
\end{figure}

In contrast, the Hybrid approach incorporates predicted future unmeasured disturbances by using a disturbance model. The key steps are as follows:  
\begin{enumerate}
\item Model identification — A gray-box building model is obtained through OD system identification.  
\item Disturbance estimation — Using the building model, unmeasured disturbances (IDs, $\hat{\zeta}_{\text{ID}}$, or ODs, $\hat{\nu}_{\text{OD}}$) are estimated via Eq.~\ref{eq:ID_filter} (ID) or Eq.~\ref{eq:od_pred} (OD).  
\item Disturbance model development — The estimated IDs or ODs are separated into input (past) and output (future) components for developing a disturbance model. For example, a graphical representation of the disturbance model for the ID case is shown in Fig.~\ref{fig:hybrid_example}.  
\item Future disturbance prediction — The disturbance model predicts future unmeasured disturbances, which are then included in the gray-box building model for future predictions. Specifically, the predicted future IDs or ODs are directly used for temperature forecasting, as shown in Eqs.~\ref{eq:ID_hybrid} and \ref{eq:od_hybrid}.  
\end{enumerate}

\begin{eqnarray}
\label{eq:ID_hybrid}
\begin{aligned}
\hat{\textbf{x}}_{\text{ID}}(k+1;\theta)&=\textbf{A}_{\text{d},\text{ID}}(\theta)\hat{\textbf{x}}_{\text{ID}}(k;\theta)+\textbf{B}_{\text{w,d}}(\theta)\textbf{w}(k)+\textbf{B}_{\text{u,d}}(\theta)\textbf{u}(k)\\
\hat{y}_\text{ID}(k;\theta)&=\textbf{C}_{\text{d},\text{ID}}\hat{\textbf{x}}_{\text{ID}}(k)\\
\end{aligned}
\end{eqnarray}
where $\hat{\textbf{x}}_{\text{ID}}(k;\theta)=\left[\hat{\textbf{x}}(k;\theta),\, \hat{\zeta}_\text{ID}(k) \right]$, and $\hat{\zeta}_\text{ID}(k)$ is predicted ID at time $k$.
\begin{eqnarray}
\label{eq:od_hybrid}
\begin{aligned}
\hat{\textbf{x}}(k+1;\theta)&=\textbf{A}_\text{d}(\theta)\hat{\textbf{x}}(k;\theta)+\textbf{B}_\text{w,d}(\theta)\textbf{w}(k)+\textbf{B}_\text{u,d}(\theta)\textbf{u}(k)\\
\hat{y}_\text{OD}(k)&=\textbf{C}_\text{d}\hat{\textbf{x}}(k;\theta)+\hat{v}_{\text{OD}}(k)
\end{aligned}
\end{eqnarray}
where $\hat{v}_{\text{OD}}(k)$ is predicted OD at time $k$.
\subsection{Design of Hybrid model structure}
\label{sec:body32}

The first step in designing the Hybrid model is to understand the relationship between the quality of the gray-box model and the estimated unmeasured disturbances obtained from it (via Eq.~\ref{eq:ID_hybrid} or \ref{eq:od_hybrid}). True unmeasured disturbances include various factors such as internal heat gains from occupants, appliances, plug loads, infiltration, and ventilation due to window openings. Because of their stochastic nature, it is common to represent the average profiles of unmeasured disturbances using time-related features (e.g., daily or weekly occupancy schedules) \cite{Ellis2021-om,Hong2020-iv}. However, when significant unmeasured disturbances are present, the accuracy of the gray-box model decreases (Section~\ref{sec:body23}). As a result, the estimated unmeasured disturbances may deviate from the true values. Therefore, it is important to understand how the quality of the gray-box model affects both the estimation of unmeasured disturbances and the overall performance of the Hybrid model.  

\begin{figure}
    \centering 
    \includegraphics[width=\textwidth]{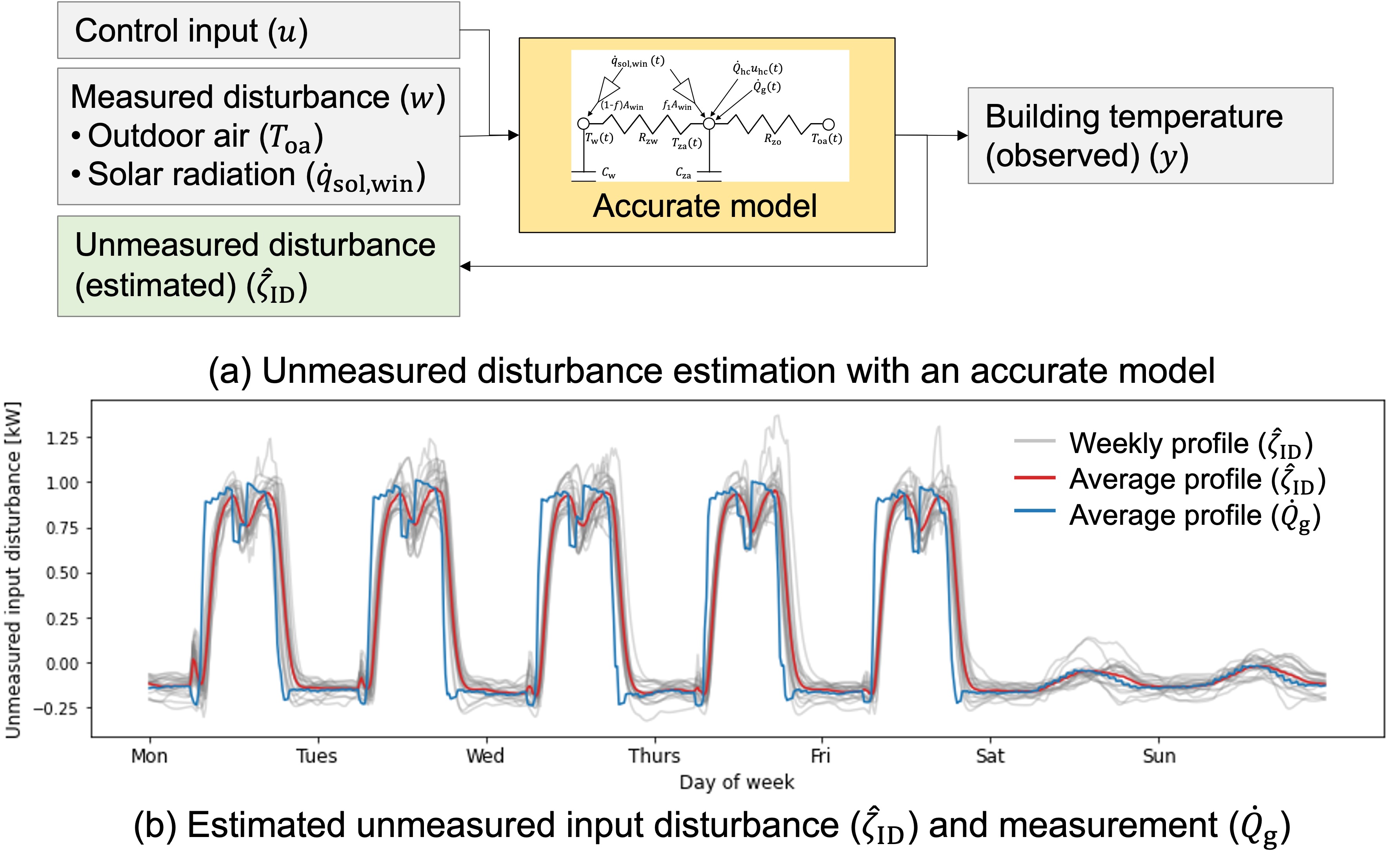} 
    \caption{Unmeasured disturbance estimation with the accurate model.} 
    \label{fig:accurate} 
\end{figure}

Figure~\ref{fig:accurate}(a) presents a schematic diagram of unmeasured disturbance estimation ($\hat{\zeta}_{\text{ID}}$) using an accurate gray-box (TRUE) model. Based on the data shown in Fig.~\ref{fig:sysid_data}, the average estimated weekly profiles are compared with the average measured values ($\dot{Q}_{\text{g}}$) in Fig.~\ref{fig:accurate}(b). As shown in Fig.~\ref{fig:sysid_data}, each day’s internal heat gains were generated according to human behavior patterns with added random noise. Consequently, the estimated weekly profiles also exhibit a recurring weekly pattern with some variability, and their average profile ($\hat{\zeta}_\text{ID}$) closely matches the average measured profile ($\dot{Q}_{\text{g}}$). This demonstrates that when the gray-box model is accurate, it is possible to extract meaningful information about unmeasured disturbance profiles from the data, which can then be modeled using time-related features.  

\begin{figure}
    \centering 
    \includegraphics[width=\textwidth]{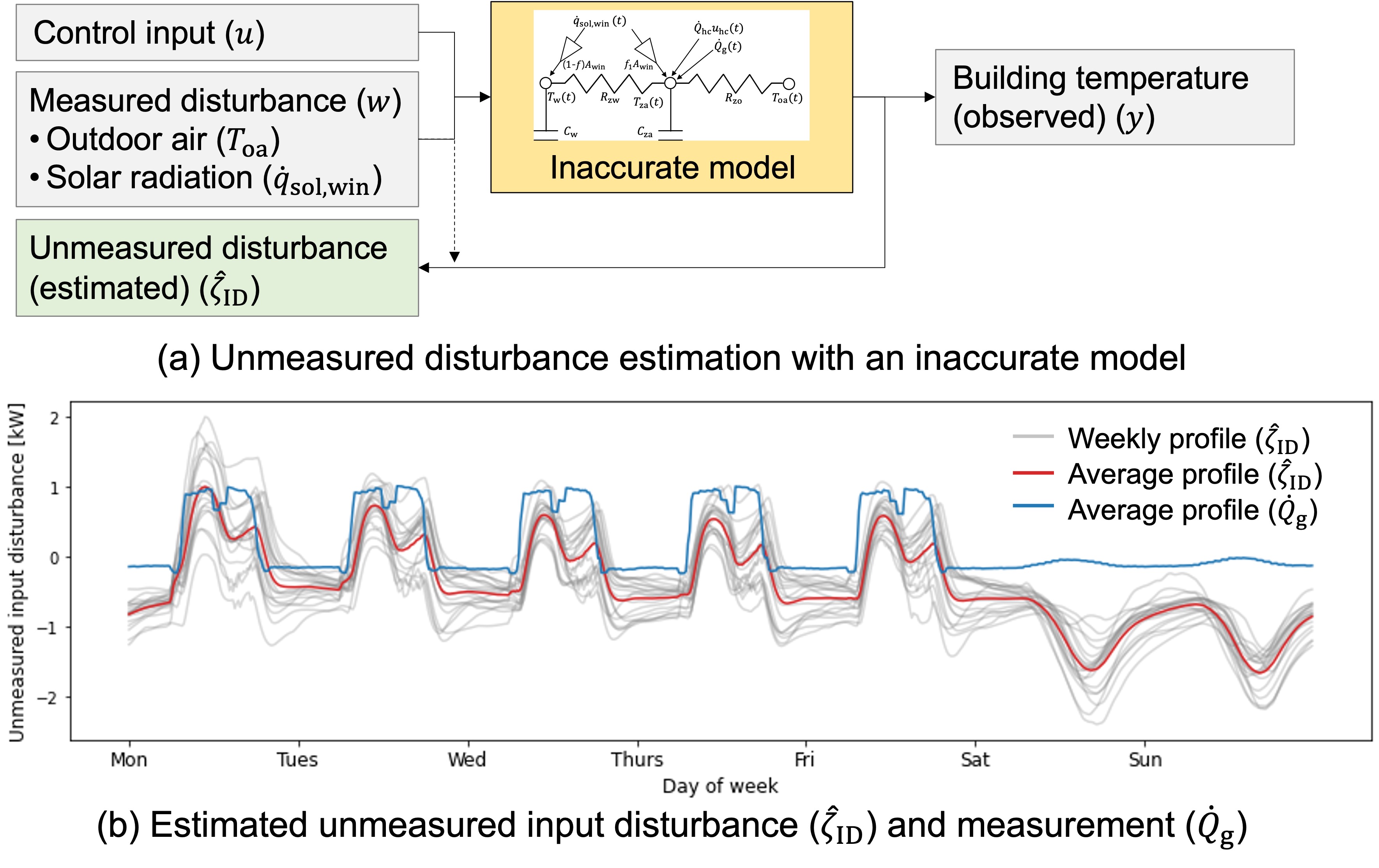} 
    \caption{Unmeasured disturbance estimation with the inaccurate model.} 
    \label{fig:inaccurate} 
\end{figure}

However, when the estimated gray-box model is inaccurate due to significant unmeasured disturbances, the estimated unmeasured disturbances do not reliably represent the true disturbance profiles. Figure~\ref{fig:inaccurate}(a) shows a schematic diagram illustrating the estimation of unmeasured disturbances using an inaccurate model, while Fig.~\ref{fig:inaccurate}(b) compares the average estimated weekly profiles ($\hat{\zeta}_{\text{ID}}$) with the measured values ($\dot{Q}_{\text{g}}$). Although the estimated disturbances still exhibit weekly patterns, they deviate from the measured values. Furthermore, compared to the accurate model case shown in Fig.~\ref{fig:accurate}(a), the week-to-week estimations display greater variability, suggesting the presence of additional dynamics beyond internal heat gains. Notably, as illustrated in Fig.~\ref{fig:inaccurate}(a), information from the control inputs and measured disturbances must be incorporated into the estimation process when the system model lacks accuracy.

In this research, a deep learning model was selected to represent unmeasured disturbances among various machine-learning approaches. Deep learning has emerged as one of the most successful machine learning techniques in recent years, owing to its flexibility and ability to capture the nonlinear and complex characteristics of data, supported by rapid advancements in computing power \cite{Murphy2022-zn}. Consequently, deep learning models have been widely applied in time-series forecasting \cite{Ismail_Fawaz2019-jd,Lara-Benitez2021-sh,Yazici2022-vw}. Moreover, deep learning models can inherently account for characteristics of time-series data—such as stationarity, seasonality, and dynamic structure—without the need to explicitly define them in the forecasting model \cite{Nielsen2019-hm}.  

Due to the nature of deep-learning model, it involves an infinite number of possible hyperparameter combinations (e.g., network size, activation functions, optimization parameters). The complexity of a deep learning model is generally governed by four factors: model framework, model size, optimization process, and data complexity \cite{Hu2021-rj}. Model framework refers to the type of architecture used (e.g., feedforward neural network, dynamic neural network) and the choice of activation functions. Model size relates to the number and width of hidden layers. The optimization process refers to the settings of the optimizer, such as the optimization algorithm and learning rate. Finally, data complexity concerns the dimensionality, distribution, and volume of the data.  

\begin{figure}
    \centering 
    \includegraphics[width=\textwidth]{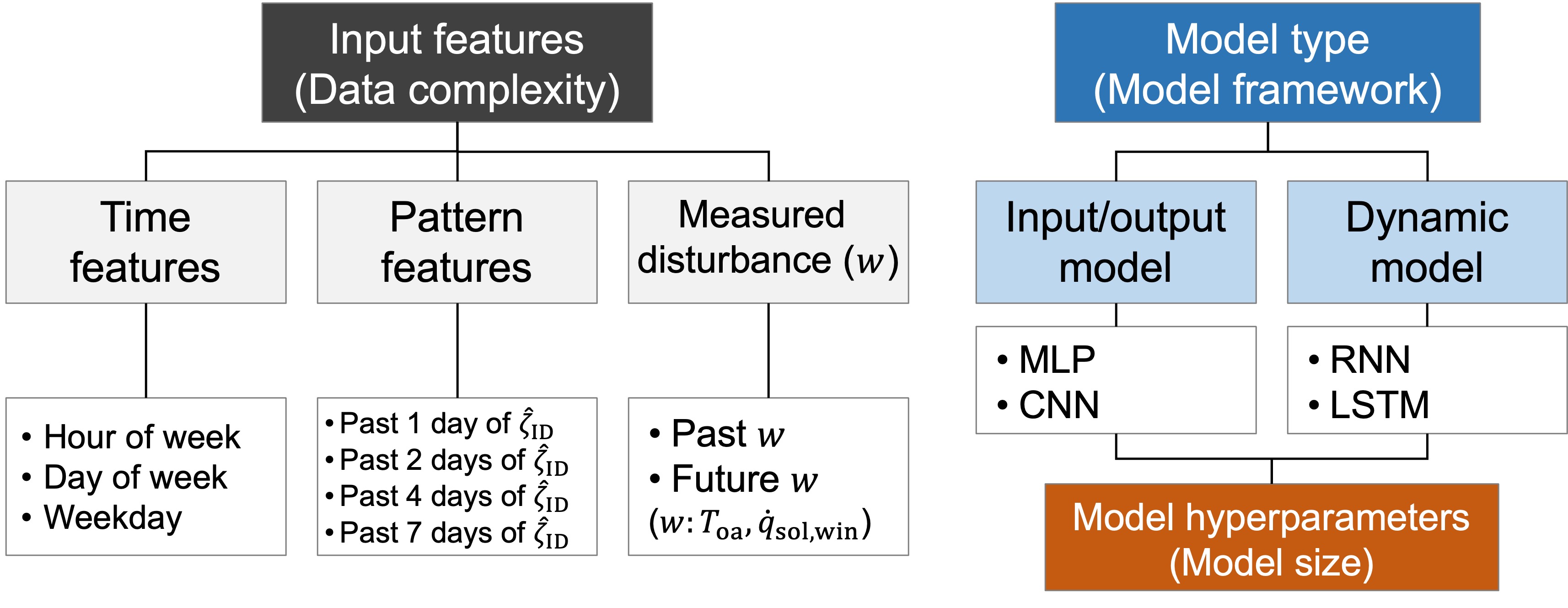} 
    \caption{Model design matrix for the unmeasured disturbance model.} 
    \label{fig:design_matrix} 
\end{figure}

Unmeasured disturbance data typically exhibits site-specific stochastic profiles, but often with periodic patterns (i.e., different buildings have different profiles). Therefore, it is important to identify model types and hyperparameters that can provide robust performance across different situations (i.e., buildings) by effectively capturing periodic patterns in the data. To explore various combinations, a model design matrix for the deep learning model (Fig.~\ref{fig:design_matrix}) was created based on the factors influencing model complexity, with the goal of selecting the most appropriate architecture. The optimization process factor was omitted because, for this relatively low-order problem (i.e., weekly patterns), it is not difficult to reach a global minimum without modifying optimization hyperparameters (e.g., learning rate).  

The input features, which determine the data dimensionality, consist of three components: time, pattern, and measured disturbance. Time features (e.g., hour of the week) and pattern features (past $n$ days of $\hat{\zeta}_{\text{ID}}$) are commonly used in time-series modeling, as they capture time-specific and autoregressive characteristics. In addition, measured disturbances ($w$), as shown in Fig.~\ref{fig:inaccurate}, are also included as input features.  

Four types of models are investigated in this study: Multi-Layer Perceptron (MLP), Convolutional Neural Network (CNN), Recurrent Neural Network (RNN), and Long Short-Term Memory (LSTM). The MLP and CNN have an input–output structure that maps an input time-series vector ($\boldsymbol{\psi} \in \mathbb{R}^{(n_{\psi}n_{\text{k},\psi})}$) to an output time-series vector ($\boldsymbol{\xi} \in \mathbb{R}^{n_{\text{k},\xi}}$), referred to as a feed-forward network. The input time-series vector is constructed by concatenating all input features over the input time period ($n_{\text{k},\psi}$) into a one-dimensional vector. The output time-series vector is a one-dimensional vector representing the estimated unmeasured disturbance for the prediction time horizon ($n_{\text{k},\xi}$).  

In contrast, the RNN and LSTM are dynamical models that map the current input features ($\boldsymbol{\psi}(k) \in \mathbb{R}^{n_\psi}$) and the previous hidden states ($\mathbf{h}(k-1)$) to the updated hidden states ($\mathbf{h}(k)$) and output ($\boldsymbol{\xi}(k) \in \mathbb{R}^{1}$) at each time step $k$—a structure referred to as a feedback network.  

Specifically, the MLP is a fully connected feed-forward neural network \cite{Murphy2022-zn}. In each layer, the input time-series vector ($\boldsymbol{\psi}$) passes through a linear transformation (defined by weights $\mathbf{W}_i$ and biases $\mathbf{b}_i$), followed by an activation function ($\varphi$) and a dropout layer. This process is repeated $n_\text{layer}$ times. Finally, an additional linear transformation (with parameters $\mathbf{W}_0$ and $\mathbf{b}_0$) is applied to produce the output ($\boldsymbol{\xi}$), as expressed in Eq.~\ref{eq:MLP}.  

\begin{eqnarray}
\begin{aligned}
\label{eq:MLP}
&\textbf{z}_1=\text{dropout}(\varphi(\textbf{W}_1 \boldsymbol{\psi}+\textbf{b}_1))\\
&\text{for }\,  i \text{ in } 2:n_\text{layer}:\\
&\,\,\,\,\,\,\,\,\,\textbf{z}_{i}=\text{dropout}(\varphi(\textbf{W}_{i} \textbf{z}_{i-1}+\textbf{b}_i))\\
&\boldsymbol{\xi}=\textbf{W}_\xi \textbf{z}_{n_\text{layer}}+\textbf{b}_{\xi}
\end{aligned}
\end{eqnarray}
where $\textbf{W}_{i}$ and $\textbf{b}_{i}$ are weights and bias parameters that maps hidden layers from  $\textbf{z}_{i-1}$ to  $\textbf{z}_{i}$, $\textbf{z}_{i}$ $\in \mathbb{R}^{n_\text{z}}$  is hidden variable of $i$ layer. $\varphi$ is activation function, $\text{dropout}$ is a dropout layer, which is commonly used for preventing over-fitting \cite{Murphy2022-zn}, and the following model size combinations were investigated ($n_\text{layer}\in\left[1,2,4\right]$, $n_\text{z}\in\left[10,20,50,100\right]$, and $\varphi\in\left[\text{ReLU},\text{SELU},\text{GELU}\right]$ \cite{Murphy2022-zn}).

CNN is widely used for image classification problems because it automatically extracts meaningful features from raw data through the convolution kernel \cite{Murphy2022-zn}. It has a similar structure to MLP, but the convolution filter and max-pooling layer are used instead of the linear layer, and one more linear layer is for the last convolution layer (Eq. \ref{eq:CNN}).
\begin{eqnarray}
\begin{aligned}
\label{eq:CNN}
&\textbf{z}_0=\varphi(\text{maxpool}(\text{conv}(\boldsymbol{\psi})))\\
&\text{for }\,  i \text{ in } 1:n_\text{layer}:\\
&\,\,\,\,\,\,\,\,\,\textbf{z}_i=\varphi(\text{maxpool}(\text{conv}(\textbf{z}_{i-1})\\
&\boldsymbol{\xi}=\textbf{W}_\xi(\text{dropout}(\varphi(\textbf{W}_{n_\text{layer}+1} \textbf{z}_{n_\text{layer}}+\textbf{b}_{n_\text{layer}+1})))+\textbf{b}_\xi
\end{aligned}
\end{eqnarray}
where $\text{conv}$ is a convolution filter and $\text{maxpool}$ is a max-pooling layer, and other notations are the same as MLP. The following model size combinations were investigated ($n_\text{layer}\in\left[1,2\right]$, $n_\text{z}\in\left[10,50,100\right]$, $n_\text{channel}\in\left[10,50,100\right]$, $n_\text{filter}\in\left[6,12\right]$, $n_\text{pool}\in\left[0,4\right]$, and $\varphi\in\left[\text{ReLU},\text{SELU},\text{GELU}\right]$ \cite{Murphy2022-zn}).

RNN is a neural network that uses an RNN cell \cite{Murphy2022-zn,PyTorch_Contributors2023-xi} as a recurrent layer. The cell maps input ($\boldsymbol{\psi}(k)\in \mathbb{R}^{n_\psi}$) and previous hidden states ($\textbf{h}(k-1)$) to the current hidden states ($\textbf{h}(k)$). The output ($\boldsymbol{\xi}(k) \in \mathbb{R}^{1}$) is calculated by applying a linear ($\textbf{W}_h$ and $\textbf{b}_h$), an activation, a dropout, and a linear ($\textbf{W}_\xi$ and $\textbf{b}_\xi$) layers to the hidden states ($\textbf{h}(k)$) for each time step ($k$) (Eq. \ref{eq:RNN}).
\begin{eqnarray}
\begin{aligned}
\label{eq:RNN}
&\text{for }\,  k \text{ in } 1:(n_{\text{k},\psi}+n_{\text{k},\xi}):\\
&\,\,\,\,\,\,\,\,\,\textbf{h}(k)=\text{RNN}
\left(\textbf{h}(k-1),\boldsymbol{\psi}(k-1)\right)\\
&\text{for }\,  k \text{ in } 1:(n_{k,\xi}):\\
&\,\,\,\,\,\,\,\,\,\boldsymbol{\xi}(k)=\textbf{W}_{\xi}(\text{dropout}(\varphi(\textbf{W}_{h} \textbf{h}(k)+\textbf{b}_{h})))+\textbf{b}_\xi\\
\end{aligned}
\end{eqnarray}
where $\textbf{h}$ is a hidden state vector that holds memory. $\boldsymbol{\psi}(k)$ is subset of $\left(\text{time}(k),w(k),\hat{\zeta}_{\text{ID}}(k)\right)$, and $\hat{\xi}_{(n_{\text{k},\psi}+1):(n_{\text{k},\psi}+n_{\text{k},\xi})}$ are used to for $\hat{\zeta}_{(n_{\text{k},\psi}+1):(n_{\text{k},\psi}+n_{\text{k},\xi})}$. The following model size combinations were investigated ($n_\text{layer}\in\left[1,2,4\right]$, $n_\text{z}\in\left[10,20,40,60\right]$, and $\varphi\in\left[\text{ReLU},\text{SELU},\text{GELU}\right]$ \cite{Murphy2022-zn}).

LSTM is one type of RNN using a LSTM cell \cite{Murphy2022-zn,PyTorch_Contributors2023-aw}. It maps input ($\boldsymbol{\psi}(k)\in \mathbb{R}^{n_\psi}$), previous hidden states ($\textbf{h}(k-1)$), and previous cell states ($\textbf{c}(k-1)$) to the current hidden ($\textbf{h}(k)$) and cell stats ($\textbf{c}(k)$). The output ($\boldsymbol{\xi}(k) \in \mathbb{R}^{1}$) is calculated by applying a linear ($\textbf{W}_h$ and $\textbf{b}_h$), an activation, a dropout, and a linear ($\textbf{W}_\xi$ and $\textbf{b}_\xi$) layers to the hidden states ($\textbf{h}(k)$) for each time step ($k$) (Eq. \ref{eq:LSTM}).
\begin{eqnarray}
\begin{aligned}
\label{eq:LSTM}
&\text{for }\,  k \text{ in } 1:(n_{k,\psi}+n_{k,\xi}):\\
&\,\,\,\,\,\,\,\,\,\textbf{h}(k),\textbf{c}(k),=\text{LSTM}
\left(\textbf{h}(k-1),\textbf{c}(k-1),\boldsymbol{\psi}(k-1)\right)\\
&\text{for }\,  k \text{ in } 1:(n_{k,\xi}):\\
&\,\,\,\,\,\,\,\,\,\boldsymbol{\xi}(k)=\textbf{W}_{\xi}(\text{dropout}(\varphi(\textbf{W}_{h} \textbf{h}(k)+\textbf{b}_{h})))+\textbf{b}_\xi\\
\end{aligned}
\end{eqnarray}
where $\textbf{h}$ and $\textbf{c}$ are hidden and cell state vector that holds short- and long-term memory, respectively. $\textbf{h}$ is a hidden state vector that holds memory. $\boldsymbol{\psi}(k)$ is subset of $\left(\text{time}(k),w(k),\hat{\zeta}_{\text{ID}}(k)\right)$, and $\hat{\xi}_{(n_{\text{k},\psi}+1):(n_{\text{k},\psi}+n_{\text{k},\xi})}$ are used to for $\hat{\zeta}_{(n_{\text{k},\psi}+1):(n_{\text{k},\psi}+n_{\text{k},\xi})}$. The following model size combinations were investigated ($n_\text{layer}\in\left[1,2,4\right]$, $n_\text{z}\in\left[10,20,40,60\right]$, and $\varphi\in\left[\text{ReLU},\text{SELU},\text{GELU}\right]$ \cite{Murphy2022-zn}).

\subsection{Model selection}
\label{sec:body33}

In this section, we present the model selection process for the unmeasured disturbance model. When selecting a deep learning model, the typical approach is to identify the optimal set of parameters that yield the best evaluation metrics while avoiding overfitting, often through techniques such as cross-validation \cite{Hastie2009-sm}, given a set of hyperparameters. Hyperparameters themselves can be optimized by repeating the parameter optimization process, for example, via Bayesian optimization \cite{Murphy2022-zn}. However, since our model design matrix is relatively small (Fig.~\ref{fig:design_matrix}), we can determine the best set of hyperparameters and parameters using a simple grid search.  

Unmeasured disturbance data is inherently site-specific (i.e., different profiles for different occupants and buildings). Therefore, it is important to identify a model that can capture the underlying pattern of the input–output relationship, as described in Section~\ref{sec:body32}. By accurately capturing this pattern, the model can achieve robust performance in extrapolation tasks without overfitting the training data. This requires first determining whether the model structure is suitable for representing the input–output relationship pattern. The expressible functional space of the model is determined primarily by the model framework (model type) and input features (data complexity). In addition, model size, defined by the depth and width of the network \cite{Hu2021-rj}, influences the model’s complexity. When the model type and input features are appropriate for capturing the input–output pattern, the model can perform well if it has at least the minimum required size (i.e., minimum model order).  

If the model is trained correctly to avoid overfitting (e.g., via cross-validation), unnecessary parameters will decay when the model size exceeds what is needed. In such cases, the model’s performance will remain consistent across a range of model sizes, indicating robustness. Conversely, if the model type and input features are poorly suited to the input–output relationship, the model will likely overfit without truly learning the underlying pattern, leading to greater sensitivity to hyperparameters related to model size.  

Our model development process follows three key steps:  

1. Model training — The dataset is split into training and testing sets. All cases in the model design matrix are included in the training phase. An optimizer identifies the parameters that minimize the training error. To prevent overfitting, we employ early stopping \cite{Murphy2022-zn}, which halts optimization when the testing error begins to increase. Each model iteration involves tuning various hyperparameters, including the number and size of hidden layers, activation functions, convolution filter dimensions, pooling layer sizes, and more. We iteratively search for the hyperparameter combination that minimizes the root-mean-squared error (RMSE) on the test set. PyTorch \cite{Paszke2019-xt} is used for model implementation and training. To assess generalizability across seasons, we train the model on one month of data and test it on the subsequent six months. We also include data from two contrasting climates: Berkeley, CA (mild climate) and Chicago, IL (four distinct seasons). All data is resampled to hourly intervals, assuming hourly occupancy patterns for unmeasured disturbances \cite{Ellis2021-om,Hong2020-iv}, and standardized using z-score normalization.  

2. Pattern expressiveness assessment — We evaluate the model’s ability to capture the input–output relationship pattern using regression-based statistical tests. Regression variables are derived from the input features and model types to assess their influence on the test error. If including certain input features or using a specific model type results in statistically significant performance improvements, we infer that the model has greater potential to capture the underlying pattern. To avoid performance degradation from undersized models, only the top three models from the regression test are considered for further evaluation. The regression model is formulated in Eq.~\ref{eq:regression}.  
\begin{eqnarray}
\begin{aligned}
\label{eq:regression}
\upsilon_{\text{test}}(c)=&\beta_{0}+\beta_{\text{CNN}}\chi_{\text{CNN}}(c)+\beta_{\text{RNN}}\chi_{\text{RNN}}+\beta_{\text{LSTM}}\chi_{\text{LSTM}}(c)+\beta_{\text{time}}\chi_{\text{time}}(c)+\\ &\beta_{\text{pattern}}\chi_{\text{pattern}}(c)+\beta_{\text{past-w}}\chi_{\text{past-w}}(c)+\beta_{\text{future-w}}\chi_{\text{future-w}}(c)+\beta_{\text{ID}}\chi_{\text{ID}}(c)
\end{aligned}
\end{eqnarray}
where $c$ is each case identifier, $\upsilon_{\text{test}}$ is a root-mean squared error of temperature prediction (Eqs. \ref{eq:ID_hybrid}-\ref{eq:od_hybrid}) on test data, $\chi_\text{CNN}$, $\chi_\text{RNN}$, and $\chi_\text{LSTM}$ are a binary indicator of whether to use each model type (i.e., 0 is MLP model), $\chi_{\text{time}}$ is a binary indicator of inclusion of any time feature, $\chi_{\text{pattern}}$ is a numeric number of the length of pattern feature, $\chi_{\text{past-w}}$ is a binary indicator of inclusion of past $w$ feature, $\chi_{\text{future-w}}$ is a binary indicator of inclusion of future $w$ feature, $\chi_{\text{ID}}$ is a binary indicator of either ID model or OD model (ID is 1).

The regression model is trained using a Bayesian approach \cite{Gelman2014-pf}, which allows direct interpretation of the marginal effect of each variable ($\beta_{*}$). For instance, if the posterior distribution of $\beta_\text{future-w}$ shows that only 1\% of samples are greater than zero, this implies that including $\beta_\text{future-w}$ in the model would reduce the prediction error with 99\% probability compared to the case without its inclusion.  

3. Once several combinations of input features and model types are identified in the second step, the robustness of these combinations is evaluated by analyzing the variation in prediction error across different model sizes. For this comparison, each pair of combinations is tested statistically to determine whether their error distributions differ significantly. Each error distribution is modeled as a log-normal distribution, and the resulting pairs are compared using the methodology described in Eqs.~\ref{eq:robustness}–\ref{eq:comparison}.  
\begin{eqnarray}
\begin{aligned}
\label{eq:robustness}
&\text{for } c \text{ in } \text{all cases} \\
&\,\,\,\,\,\,\,\text{P}(\mu_\upsilon(c))= \text{Normal}(0,10)\\
&\,\,\,\,\,\,\,\text{P}(\sigma_\upsilon(c))= \text{HalfNormal}(0,1)\\
&\,\,\,\,\,\,\,\text{P}(\upsilon_{\text{test}}(c)|\mu_\upsilon(c),\sigma_\upsilon(c))=\text{logNormal}\left( \upsilon_{\text{test}}(c)|\mu_\upsilon(c),\sigma_\upsilon(c)\right)\,\,\\
&\,\,\,\,\,\,\,\text{P}(\mu_\upsilon(c),\sigma_\upsilon(c)|\upsilon_{\text{test}}(c))\propto \text{P}(\upsilon_{\text{test}}(c)|\mu_\upsilon(c),\sigma_\upsilon(c))\text{P}(\mu_\upsilon(c))\text{P}(\sigma_\upsilon(c))\\
\end{aligned}
\end{eqnarray}
where $\mu_\upsilon(c)$ and $\sigma_\upsilon(c)$ are mean and standard deviation of Log-Normal distribution. Log-normal distribution is used because RMSE is a non-negative value.

\begin{eqnarray}
\begin{aligned}
\label{eq:comparison}
\text{Probability of }c_1\text{ is better than }c_2\text{: }\text{P}\left((\tilde{\upsilon}_{\text{test}}(c_1)|\upsilon_{\text{test}}(c_1)-\tilde{\upsilon}_{\text{test}}(c_2)|\upsilon_{\text{test}}(c_2))<0\right)\\
\end{aligned}
\end{eqnarray}
where $\tilde{\upsilon}_{\text{test}}(c)|\upsilon_{\text{test}}(c)$ is posterior predictive samples \cite{Gelman2014-pf}.

Finally, once suitable model cases that are robust with respect to model size have been identified, the final model can be selected by also considering practical factors, such as the amount of required data. For example, if the model is to be used in MPC, predictions are generated at every MPC sampling time (i.e., control time step). If the model requires $n$ days of recent data for Past $w$ and $\zeta_\text{ID}$, then $n$ days of data must be retrieved from the database for all zones at each prediction step. This can create computational and database burdens in the MPC framework. Therefore, it is advisable to minimize the amount of recent data required.  

\begin{figure}
    \centering 
    \includegraphics[width=\textwidth]{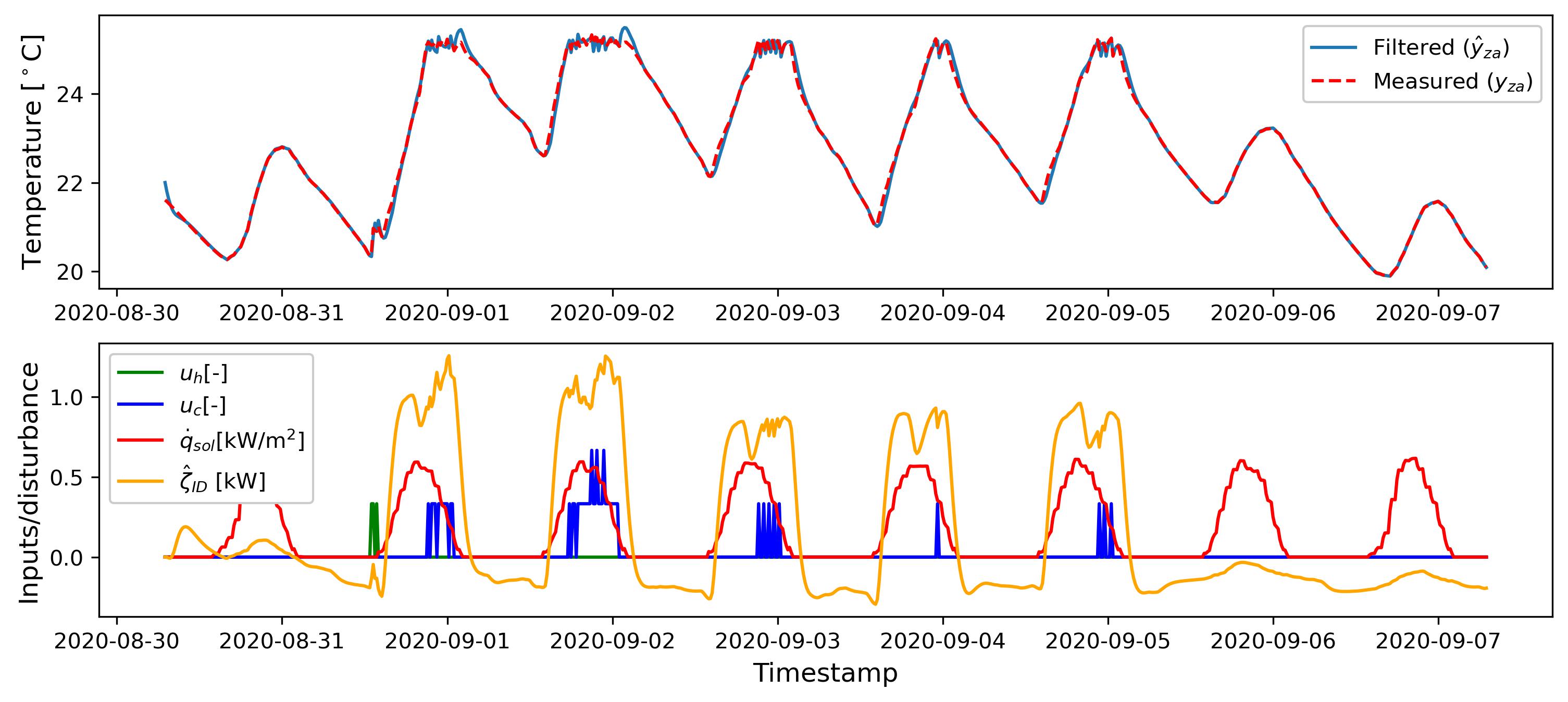} 
    \caption{Estimated input disturbance profile in a week.} 
    \label{fig:ID_profile} 
\end{figure}

\begin{figure}
    \centering 
    \includegraphics[width=\textwidth]{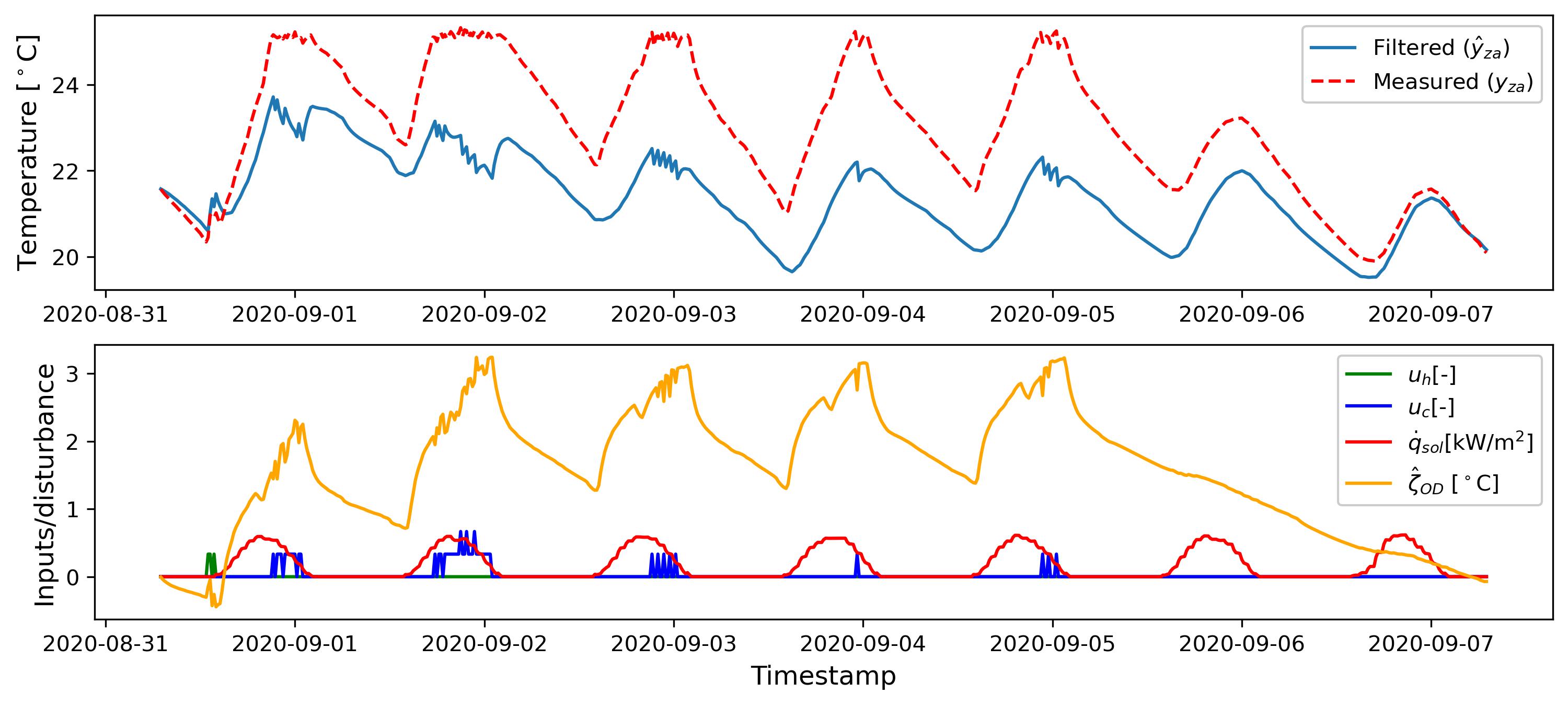} 
    \caption{Estimated output disturbance profile in a week.} 
    \label{fig:OD_profile} 
\end{figure}

In addition, the ID model may be preferred over the OD model in certain cases. Figures~\ref{fig:ID_profile} and \ref{fig:OD_profile} show the estimated ID and OD profiles for a one-week period. While the ID profile directly represents unmeasured heat gains, the OD profile reflects the cumulative impact of ID on the zone air temperature. Consequently, if the performances of the two models are similar, the ID model’s outputs are more straightforward to interpret.

\section{Result}
\label{sec:body4}

The unmeasured disturbance model was trained and evaluated using the data generated in Section~\ref{sec:body21}. However, since the weather conditions in the Berkeley, CA area are mild and lack distinct seasonal variations, additional datasets were generated to represent four seasons. These datasets were created using weather conditions from Chicago, IL \cite{US_Department_of_Energy2021-fs}, while maintaining similar building conditions.  

\subsection{Unmeasured disturbance model}
\label{sec:body41}

\begin{figure}
    \centering 
    \includegraphics[width=\textwidth]{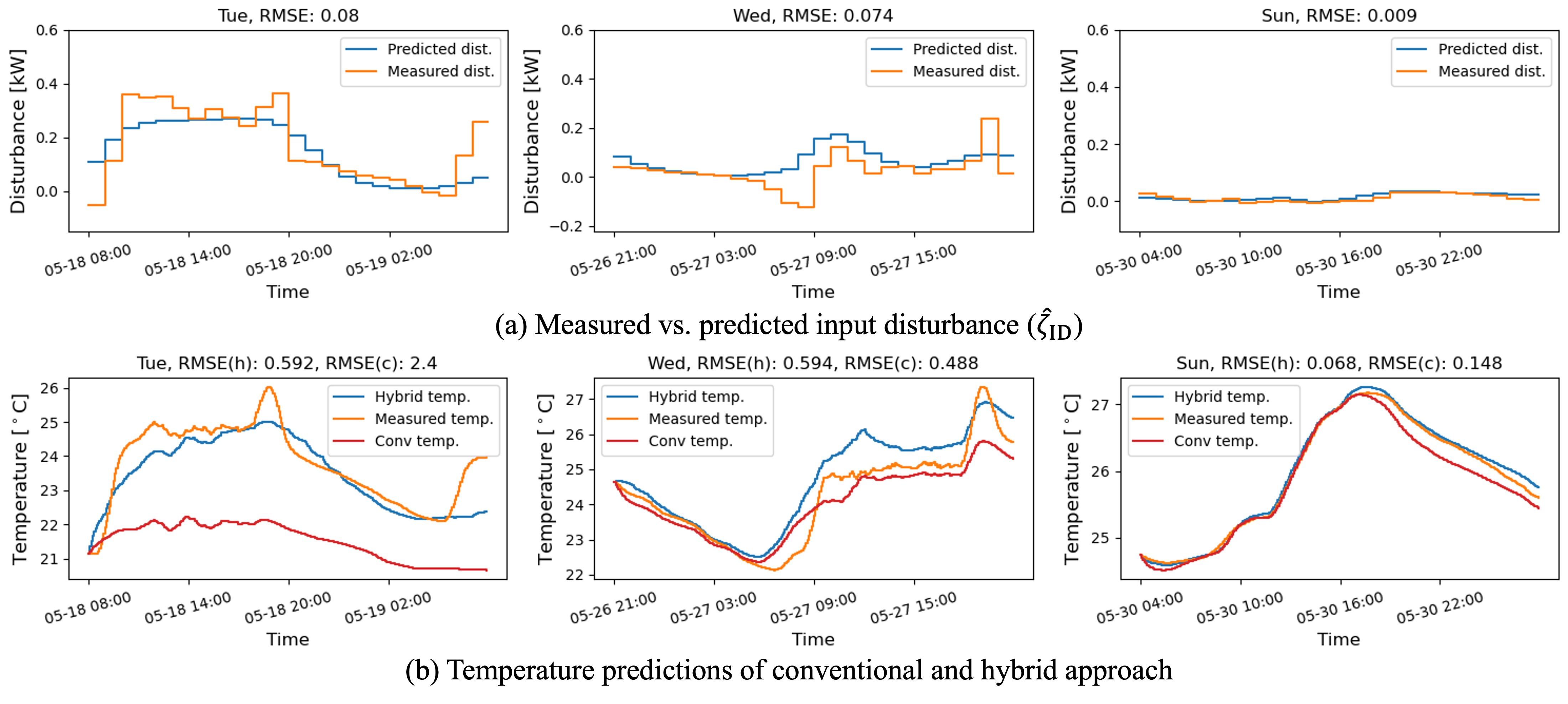} 
    \caption{Prediction results of a good model on train data; Day information and RMSE are shown on the top} 
    \label{fig:good_train} 
\end{figure}

\begin{figure}
    \centering 
    \includegraphics[width=\textwidth]{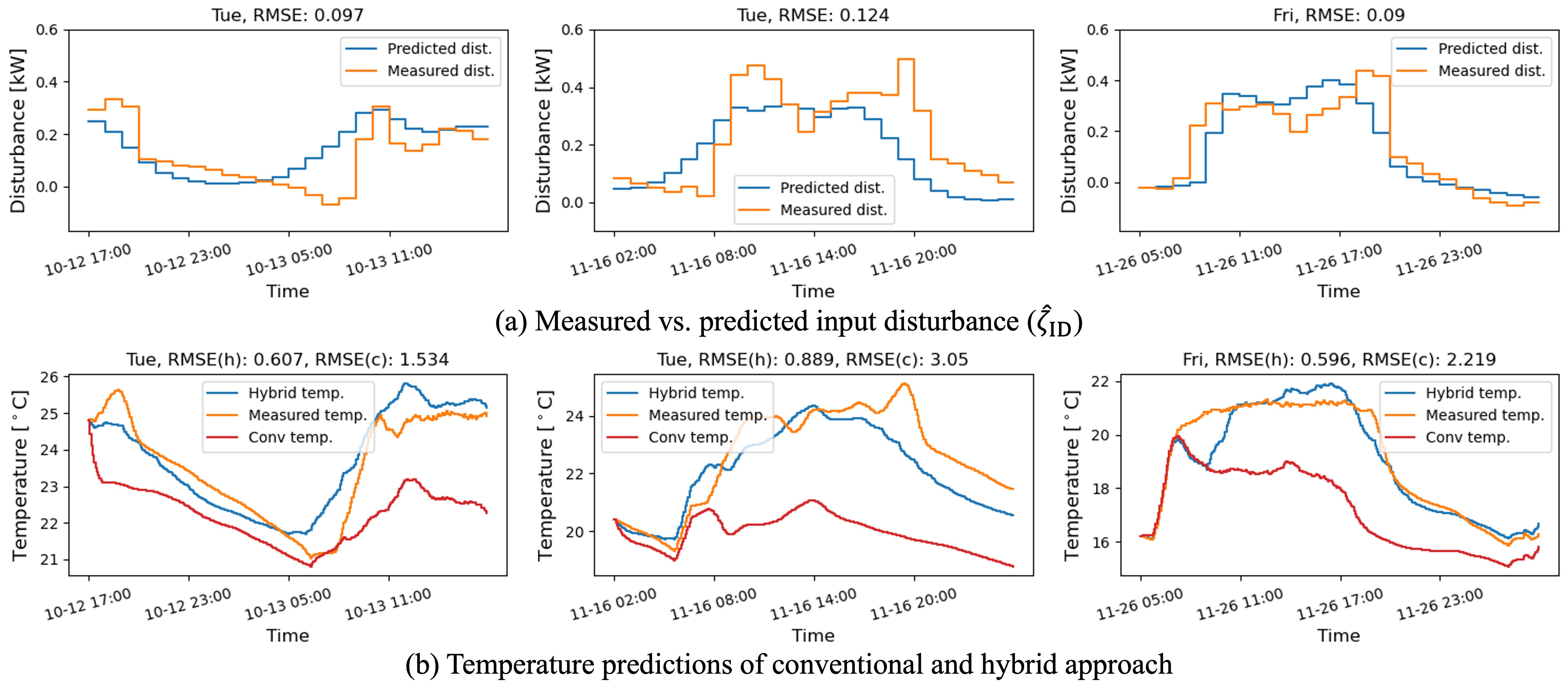} 
    \caption{Prediction results of a good model on test data; Day information and RMSE are shown on the top} 
    \label{fig:good_test} 
\end{figure}

Figs.~\ref{fig:good_train} and \ref{fig:good_test} present randomly sampled prediction results of a well-performing unmeasured disturbance (ID) model on the training and test datasets for Chicago weather, respectively. The selection of this model was carried out in the subsequent section through the model selection process. In the top panels, the predicted ID profiles are compared with the measured profiles (i.e., those estimated from data using Eq.~\ref{eq:ID_filter}). As shown in Fig.~\ref{fig:sysid_data}, the magnitude of unmeasured disturbances is generally higher on weekdays than on weekends. The selected model demonstrates strong predictive performance, with close alignment between predicted and measured profiles for both the training and test datasets. Consequently, the Hybrid approach (h) reduces the RMSE for one-day-ahead temperature prediction by approximately $0.3$–$2.0^\circ\text{C}$ compared to the Conventional approach (c).  


\begin{figure}
    \centering 
    \includegraphics[width=\textwidth]{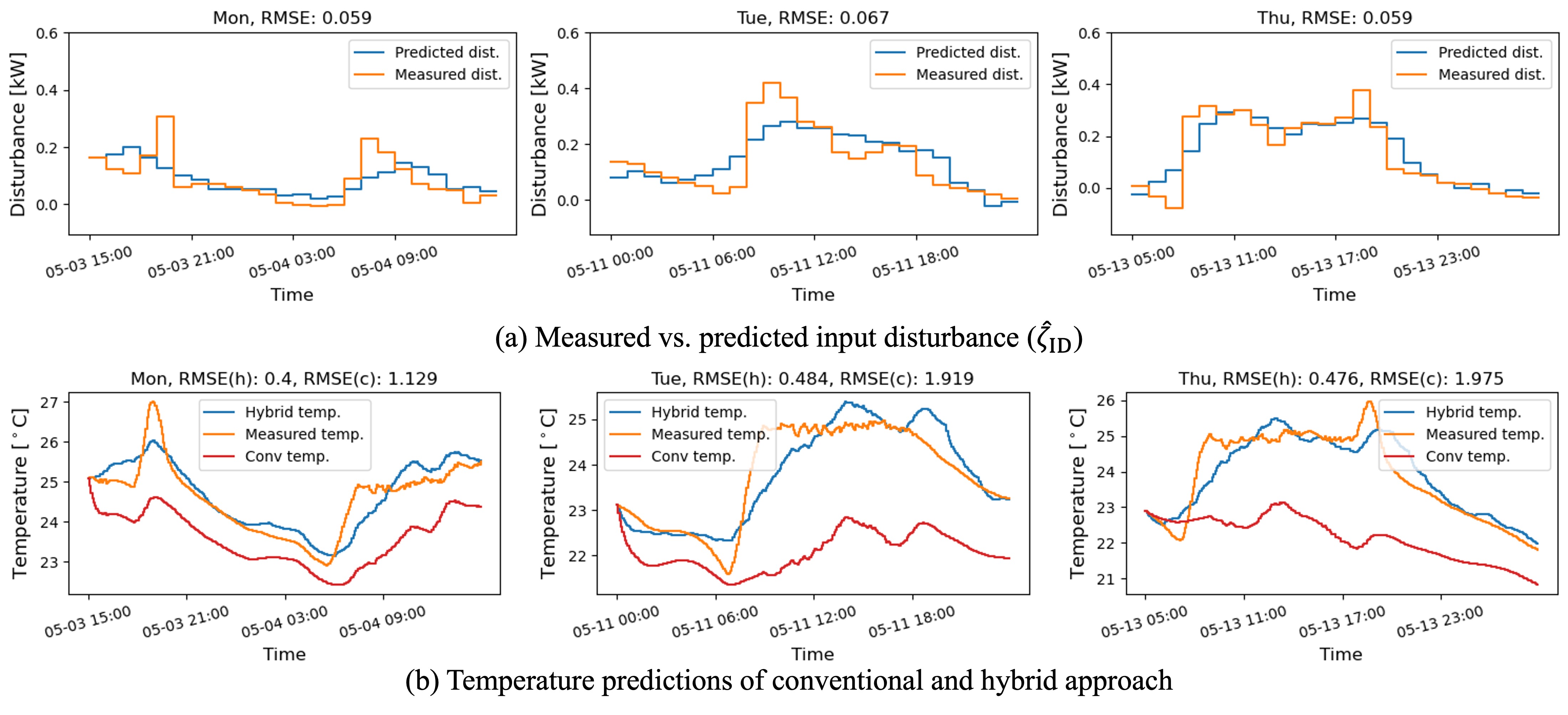} 
    \caption{Prediction results of a bad model on train data; Day information and RMSE are shown on the top} 
    \label{fig:bad_train} 
\end{figure}

\begin{figure}
    \centering 
    \includegraphics[width=\textwidth]{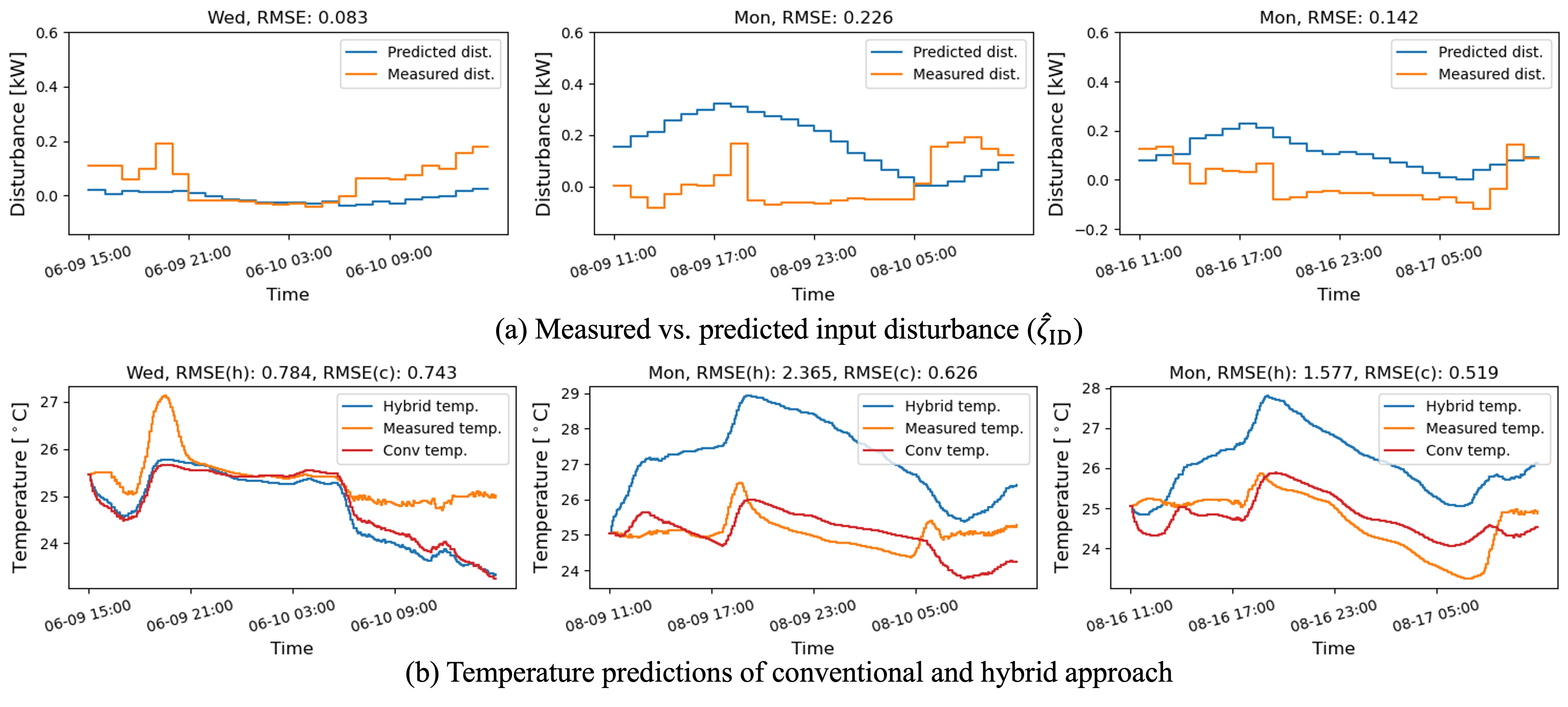} 
    \caption{Prediction results of a bad model on test data; Day information and RMSE are shown on the top} 
    \label{fig:bad_test} 
\end{figure}

Figures~\ref{fig:bad_train} and \ref{fig:bad_test} show the prediction results of a poorly performing unmeasured disturbance (ID) model on both the training and test datasets. These results underscore the critical impact of model quality on prediction performance. Although the model was trained with measures to prevent overfitting, its inadequate structure fails to capture the essential features of the data. As a result, the model exhibits overfitting and poor extrapolation capability, as evidenced by the substantial difference between its predictions on the training data (Fig.~\ref{fig:bad_train}) and the test data (Fig.~\ref{fig:bad_test}).  
\subsection{Model selection}
\label{sec:body42}

\begin{figure}
    \centering 
    \includegraphics[width=\textwidth]{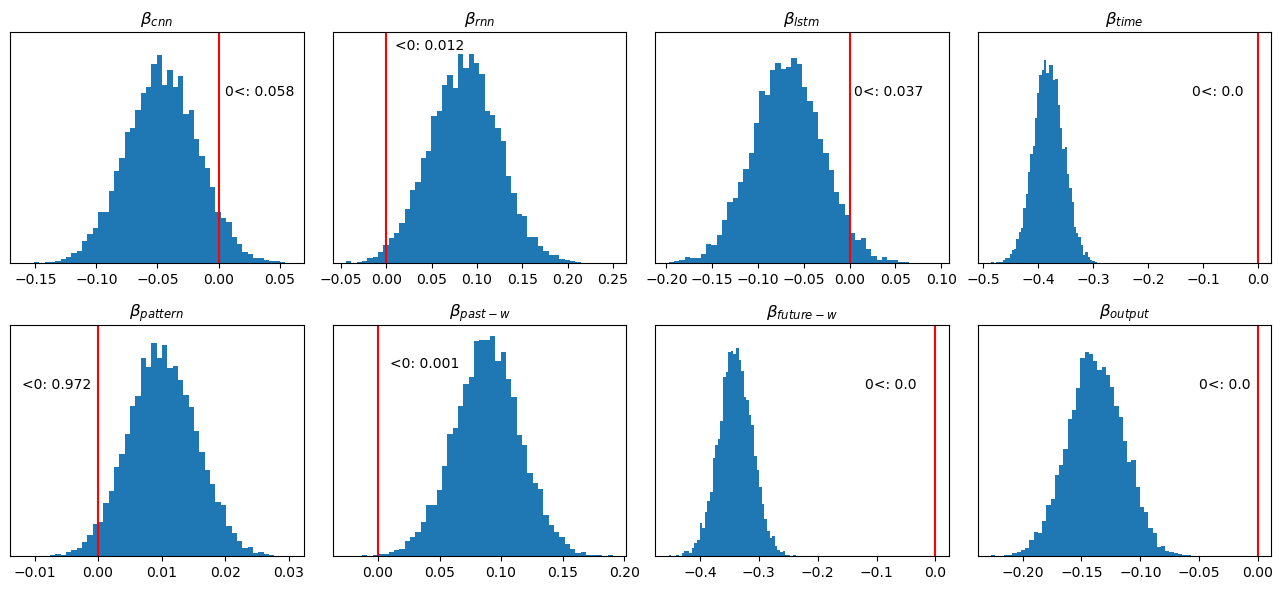} 
    \caption{Model type and input feature's marginal effect on model performance} 
    \label{fig:regression} 
\end{figure}

One of the primary contributions of this study is the demonstration of a systematic approach for determining the structure of an unmeasured disturbance model through a model selection process. As described in Section~\ref{sec:body33}, the model’s ability to capture the input–output relationship pattern is assessed using a regression model (Eq.~\ref{eq:regression}), and the posterior distributions of the regression parameters are visualized in Fig.~\ref{fig:regression}.  

Compared to the MLP model type, most CNN and LSTM samples show a reduction in prediction error, whereas the RNN does not. However, some CNN and LSTM samples exhibit errors greater than zero (opposite for the RNN). This is primarily because the regression analysis only reflects the marginal effect of each variable without accounting for potential interaction effects. Nonetheless, the inclusion of the time feature and future $w$ consistently reduces errors. In contrast, the pattern features and past $w$ tend to increase errors. Considering the scales and distributions of the regression parameters, the positive effects of the time feature and future $w$ are clear, while the effects of the other variables remain inconclusive.  

In addition, the use of the ID model ($\beta_\text{output}$) leads to lower errors. This can be attributed to the structural difference between the ID and OD models. While the OD model’s predictions are directly added to the temperature prediction, the ID model predicts unmeasured heat gains, whose influence on temperature prediction error is smaller. Based on these findings, the time feature, future $w$, and ID model are selected as suitable model components.  

To evaluate robustness, we applied Eq.~\ref{eq:robustness} to the selected regression cases. The model case with the lowest median prediction error across various model sizes was identified and compared to other cases. The comparison results, sorted by median values, are shown in Fig.~\ref{fig:robustness}, with details of each model case provided in the Appendix. Model cases that do not show statistically significant differences from the best case are highlighted in blue boxes.  

One notable observation is that the dynamical models (i.e., RNN and LSTM) exhibit narrower error distributions, indicating greater robustness to changes in model size. Since all model cases in the blue boxes show similar performance, the final model structure is selected based on practical considerations. From Fig.~\ref{fig:regression}, the LSTM is preferred over the RNN due to its stronger marginal effect. Furthermore, as discussed in Section~\ref{sec:body33}, the LSTM requires less historical data, making it more efficient in terms of database usage. Finally, we found that the hour-of-the-week (how) feature is more effective in capturing weekly patterns than the day-of-the-week (dow) or weekday features. Based on these findings, the $\texttt{DEEP\_AR\_LSTM\_FORECAST-case01}$ is chosen as the final model structure.  

\begin{figure}
    \centering 
    \includegraphics[width=\textwidth]{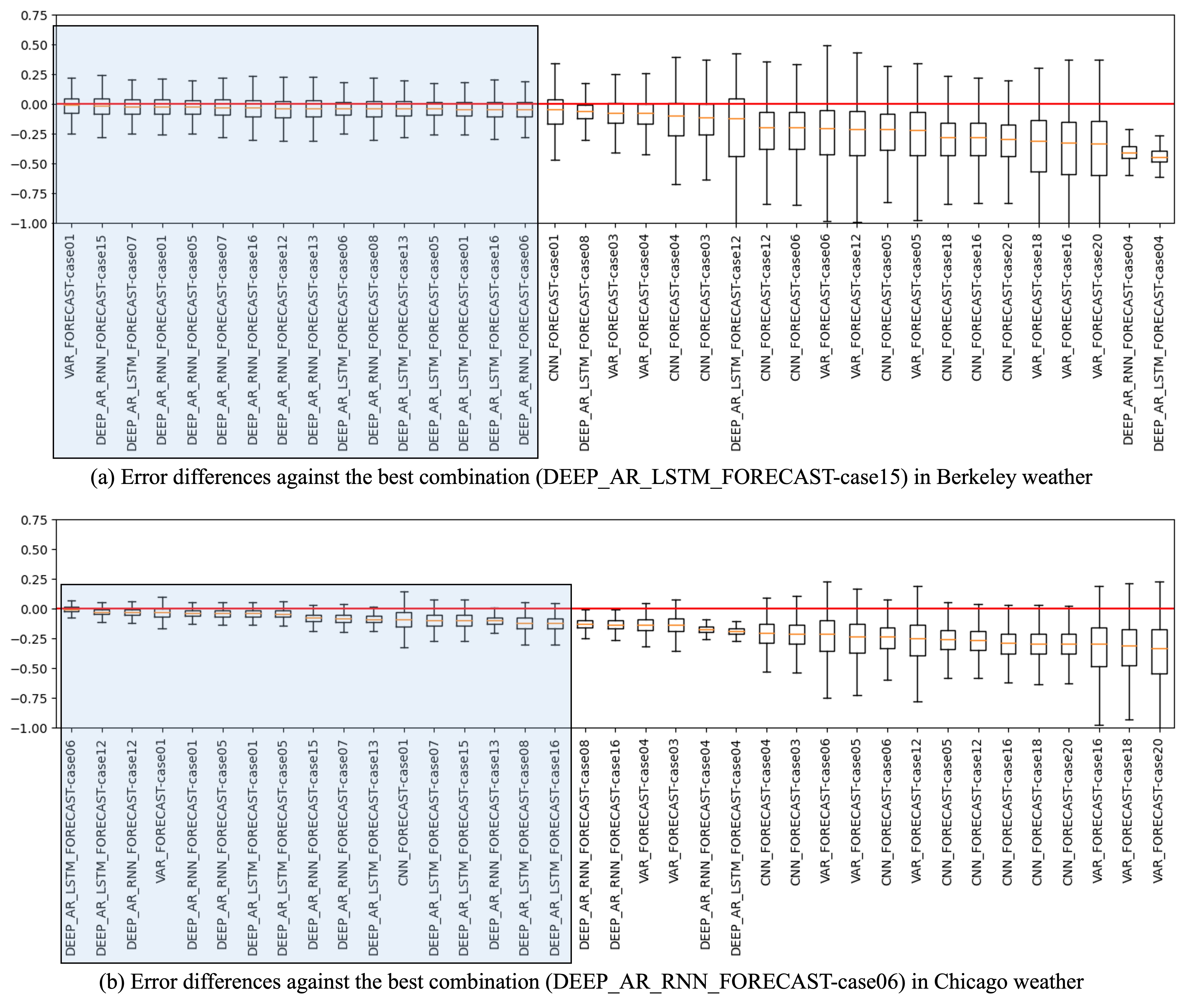} 
    \caption{Robustness of each model type and input feature combination compared to the best combination; probability lower than zero (red line) indicates the probability of best combination is better than each case} 
    \label{fig:robustness} 
\end{figure}
\begin{figure}
    \centering 
    \includegraphics[width=\textwidth]{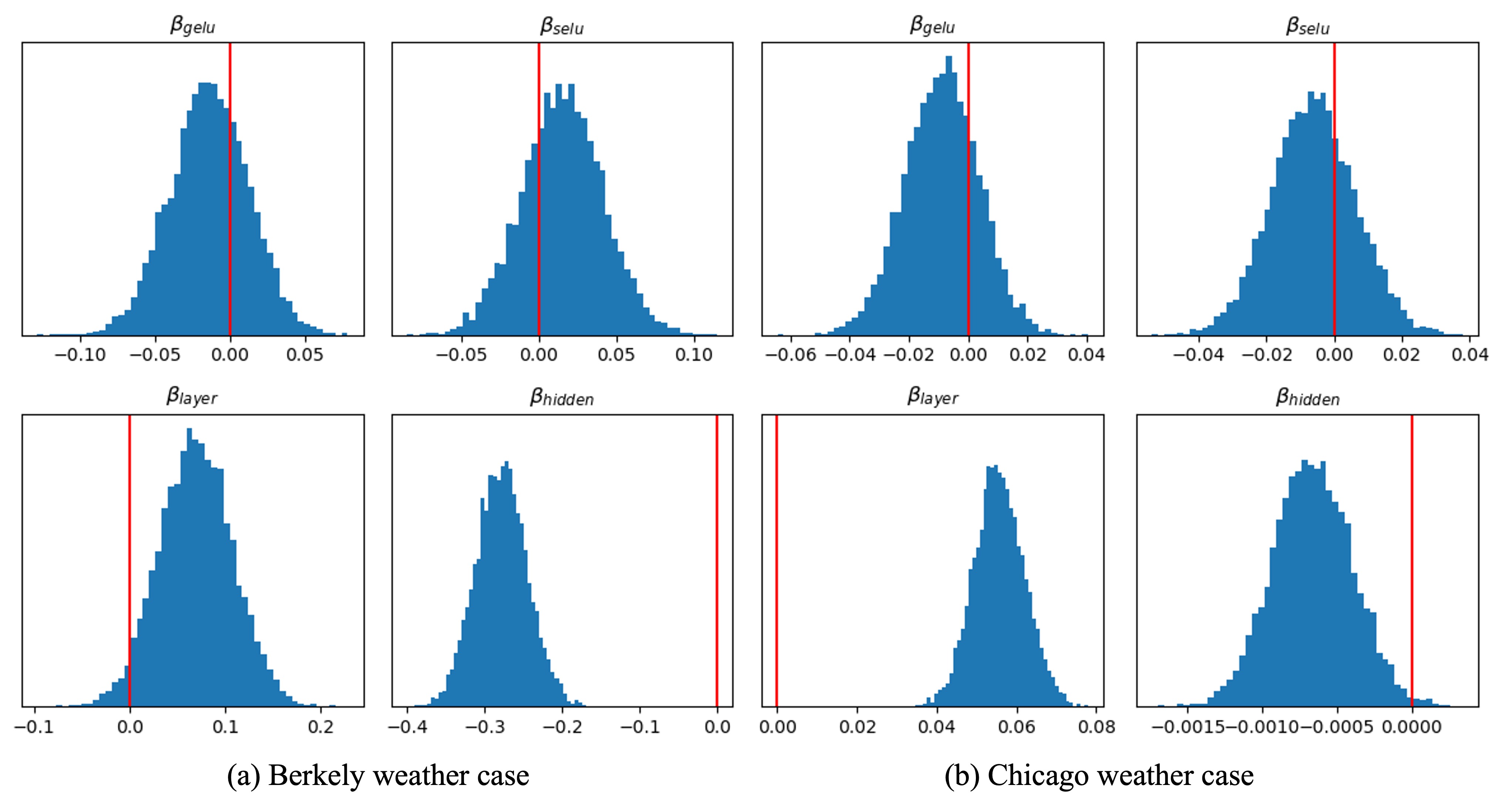} 
    \caption{Impact of model size hyperparameters on test prediction error} 
    \label{fig:model_size} 
\end{figure}

We further investigated the impact of model size hyperparameters on the test prediction error (Fig.~\ref{fig:model_size}). As described in Eq.~\ref{eq:LSTM}, we trained models with varying sizes and numbers of hidden layers, as well as different activation functions, and then applied the same regression analysis outlined in Eq.~\ref{eq:regression}. The results indicate that increasing the number of hidden layers reduces the test prediction error, whereas the other variables either had no effect or a negative effect. Based on these findings, we selected the final model configuration as $\varphi$: ReLU and $n_\text{layer}$: 1. However, the optimal size of the hidden layers is determined during the training process by selecting the configuration that yields the lowest test RMSE. 
\subsection{Prediction performance on simulated data}
\label{sec:body43}

\begin{figure}
    \centering 
    \includegraphics[width=\textwidth]{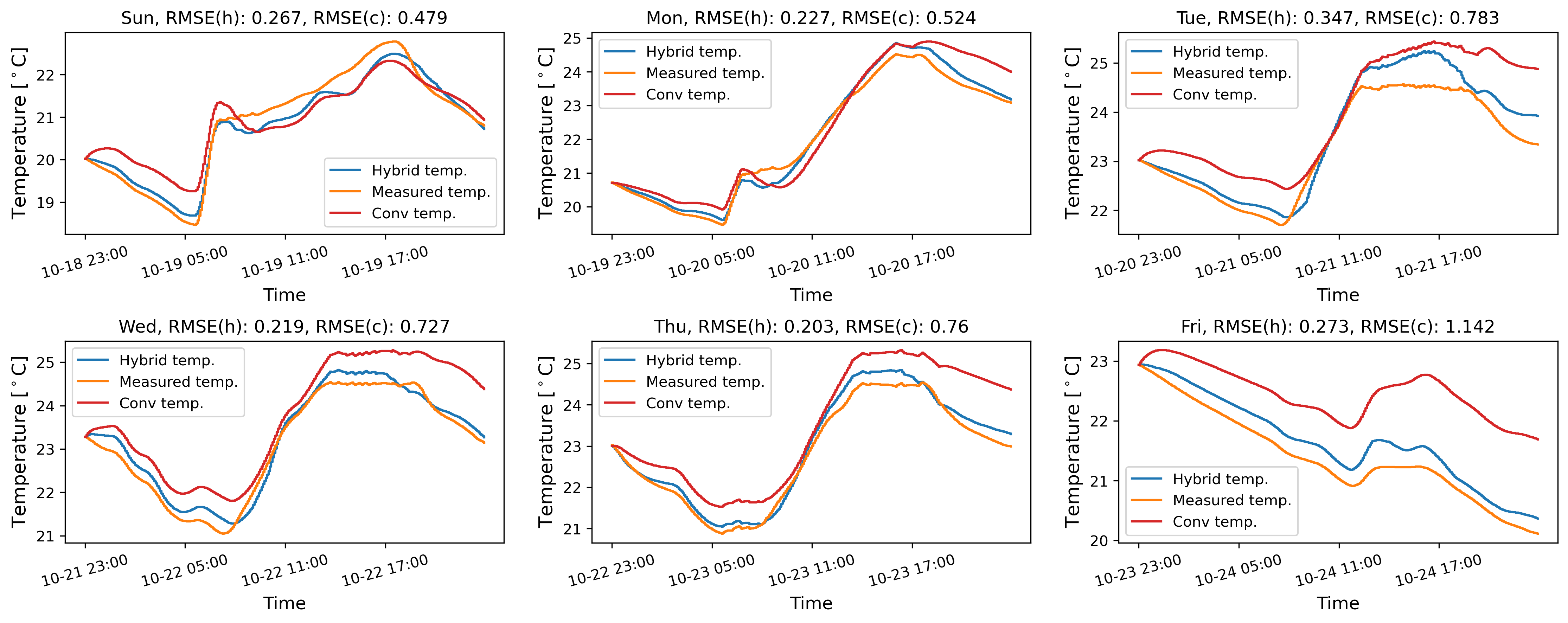} 
    \caption{Temperature predictions of Hybrid and Conventional approaches for Berkeley data} 
    \label{fig:temp_berkeley} 
\end{figure}

\begin{figure}
    \centering 
    \includegraphics[width=\textwidth]{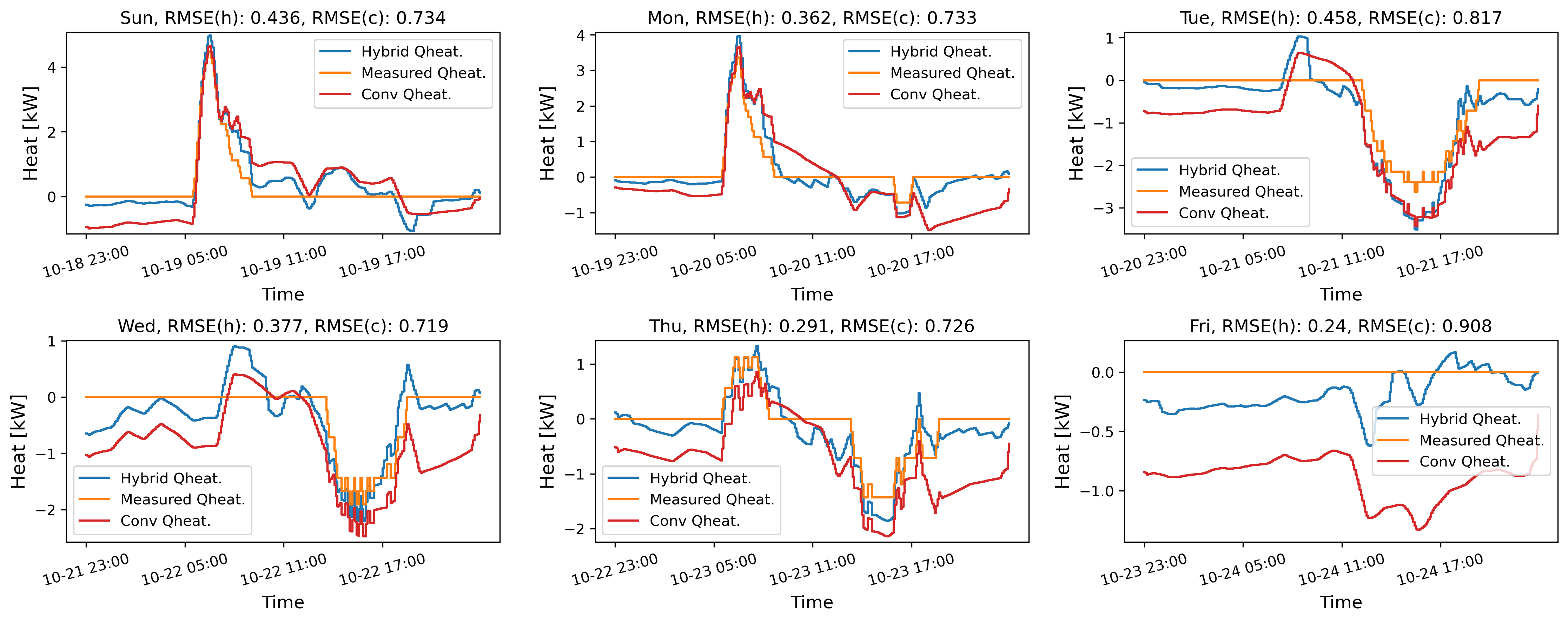} 
    \caption{Amount of required heating rate of Hybrid and Conventional approaches for Berkeley data} 
    \label{fig:Qheat_berkeley} 
\end{figure}

\begin{figure}
    \centering 
    \includegraphics[width=\textwidth]{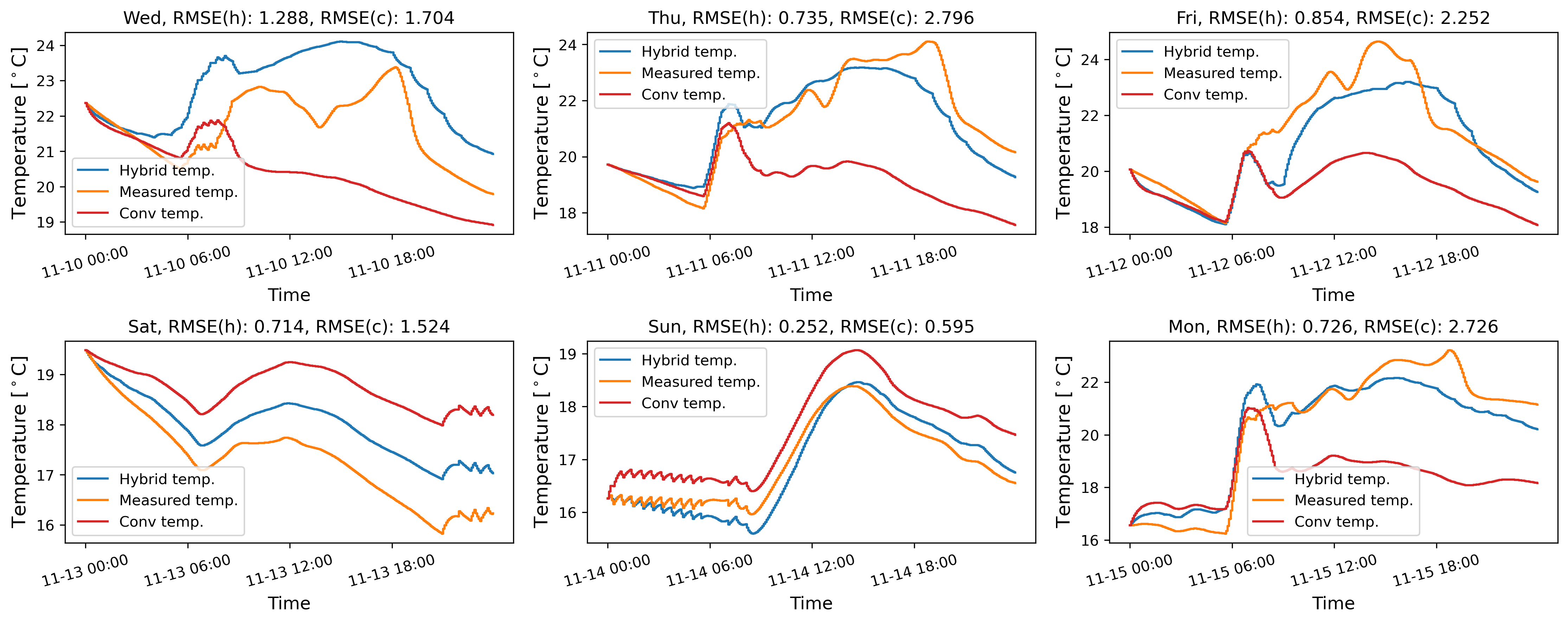} 
    \caption{Temperature predictions of Hybrid and Conventional approaches for Chicago data} 
    \label{fig:temp_chicago} 
\end{figure}

\begin{figure}
    \centering 
    \includegraphics[width=\textwidth]{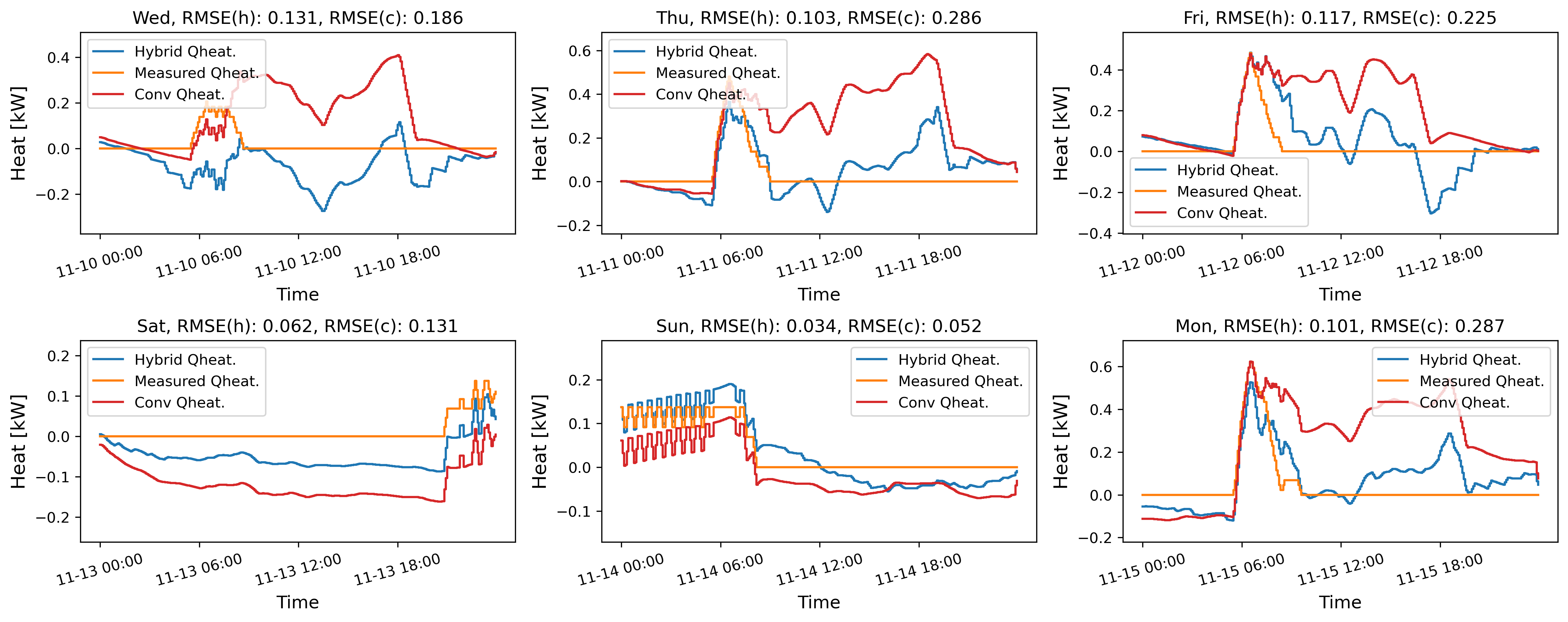} 
    \caption{Amount of required heating rate of Hybrid and Conventional approaches for Chicago data} 
    \label{fig:Qheat_chicago} 
\end{figure}

Since the models are trained using cooling season data, it is important to evaluate their prediction performance on heating season data. Figures~\ref{fig:temp_berkeley} and \ref{fig:Qheat_berkeley} present the one-day-ahead temperature predictions and the required heating load for Berkeley weather, respectively. Both the Hybrid and Conventional approaches capture the overall temperature profiles; however, the Conventional approach exhibits greater divergence over time, resulting in an RMSE increase of approximately $0.2$–$0.9^\circ$C. Similarly, both approaches capture the overall profile of the required heating load (as described in Eqs.~\ref{eq:inverseQ1}–\ref{eq:inverseQ2}), but the Hybrid approach achieves further error reduction, particularly on weekends. This improvement is attributed to the inclusion of unmeasured disturbance information in the gray-box model, as shown in Fig.~\ref{fig:nstep_woQ}.  
The temperature and required heating load predictions for the heating season in Chicago weather are shown in Figs.~\ref{fig:temp_chicago}–\ref{fig:Qheat_chicago}. Overall, both approaches perform worse than in the Berkeley case. However, the Hybrid approach achieves an RMSE reduction of approximately $0.3$–$2^\circ$C for temperature and $0.05$–$0.18$~kW for the required heating load compared to the Conventional approach.  

\subsection{Prediction performance on experimental data}
\label{sec:body44}

In this study, a Hybrid modeling approach was applied to a single-cell representation of a small office area within the research facility FLEXLAB. FLEXLAB \cite{Lawrence_Berkeley_National_Laboratory2021-ys}, located at Lawrence Berkeley National Laboratory in Berkeley, California, USA, is a well-instrumented experimental test facility specifically designed for evaluating advanced building and grid technologies. For this experiment, a packaged heat pump rooftop unit (HP-RTU) was retrofitted to test an advanced control algorithm. The control experiment results, including the application of the Hybrid approach, are presented in a separate paper \cite{Ham2024-kz}. The focus of this paper is on evaluating the prediction performance of the Hybrid modeling approach.  

\begin{figure}
    \centering 
    \includegraphics[width=\textwidth]{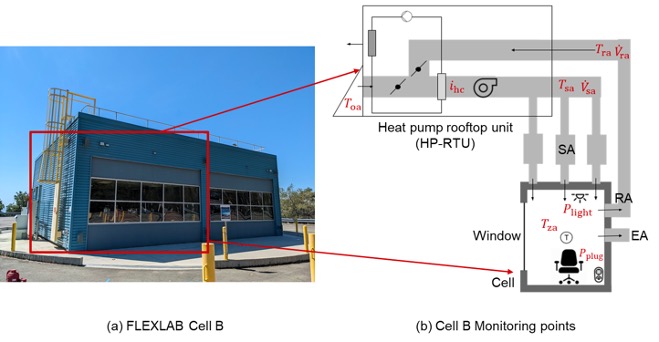} 
    \caption{FLEXLAB Cell B and its mechanical schematic diagram} 
    \label{fig:flexlab} 
\end{figure}

The experimental cell represents a small office space with a floor area of 57~m$^2$ and a large north-facing window. The HP-RTU (AAON RQ 2-ton unit) features two-stage heating and cooling control, with nominal capacities of 6.53~kW for heating and 6.16~kW for cooling at 700~CFM (1190~CMH). The HP-RTU is controlled by a Schneider Electric SE8600 thermostat, with heating and cooling stages determined by the thermostat’s internal control logic. Conditioned air is delivered through ceiling-mounted supply and return grilles via ducts connected to the HP-RTU. Although there is no dedicated exhaust fan, air is naturally exhausted through an exhaust grille connected to the ceiling plenum. Monitoring points are indicated in red text in Fig.~\ref{fig:flexlab}(b).  

%%%%%%%% start from here %%%%%%%%%%%%%%%%

The occupied setpoint ranged from 70$^\circ$F (21.1$^\circ$C) for heating to 74$^\circ$F (23.3$^\circ$C) for cooling during daytime hours (6:00–18:00 Pacific Time), while the unoccupied setpoint was 60$^\circ$F (15.6$^\circ$C) for heating and 80$^\circ$F (26.7$^\circ$C) for cooling during nighttime hours (18:00–6:00 Pacific Time). The HVAC system operated on a fixed weekday schedule (Monday–Friday, with weekends following weekday settings), and climate conditions were typical for Berkeley, CA. The supply fan operated at a fixed speed of 950~CFM (1614~CMH) (85\% fan speed), with a minimum ventilation air supply of 155~CFM (263~CMH) set by a fixed outdoor air damper position. The heating supply air temperature was controlled by the thermostat, with a maximum limit of 100$^\circ$F (37.8$^\circ$C) to prevent hot air short-circuiting. The air-side economizer mode was disabled and kept at its minimum position.  

\begin{figure}
    \centering 
    \includegraphics[width=\textwidth]{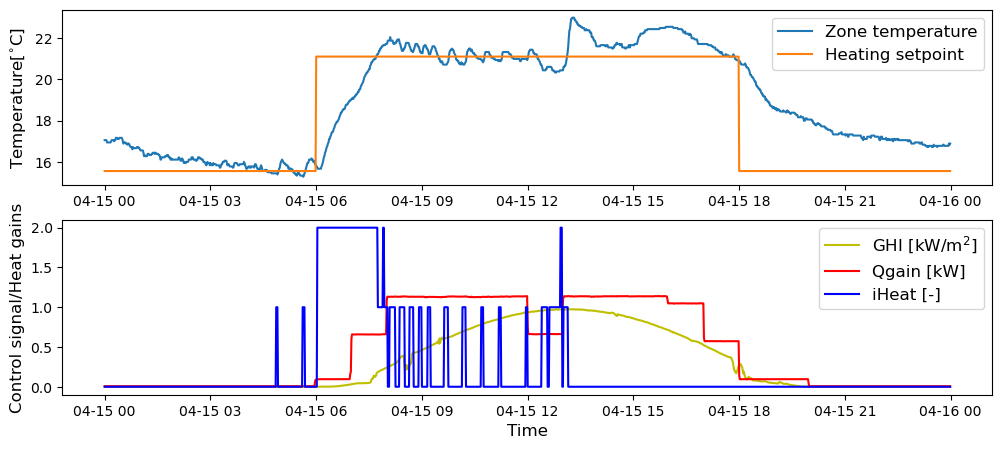} 
    \caption{A daily operational profile of the HP-RTU in the experimental cell} 
    \label{fig:flexlab_data} 
\end{figure}

Occupancy loads varied throughout the day: fully occupied periods (100\%, 600~W) occurred from 08:00–12:00 and 13:00–17:00, while partially occupied periods (50\%, 300~W) occurred from 07:00–08:00, 12:00–13:00, and 17:00–18:00. Lighting levels were adjusted accordingly, with full lighting (100\%, 350~W) from 07:00–16:00 and partial lighting (50\%, 175~W) from 06:00–07:00 and 16:00–20:00. A typical daily operational profile of the HP-RTU in the experimental cell is shown in Fig.~\ref{fig:flexlab_data}.  

After applying setpoint perturbations (Section~\ref{sec:body22}), system identification was performed using the OD approach described in Section~\ref{sec:body223}. Subsequently, one month of data was used to train the unmeasured disturbance model. The final model, $\texttt{DEEP\_AR\_LSTM\_FORECAST-case01}$, selected in Section~\ref{sec:body42}, was trained with a maximum of 1000 epochs and an early stopping patience of 150 epochs. Training for each feature set was repeated 10 times, and the configuration yielding the lowest mean squared error on one week of test data was selected.  

\begin{figure}
    \centering 
    \includegraphics[width=\textwidth]{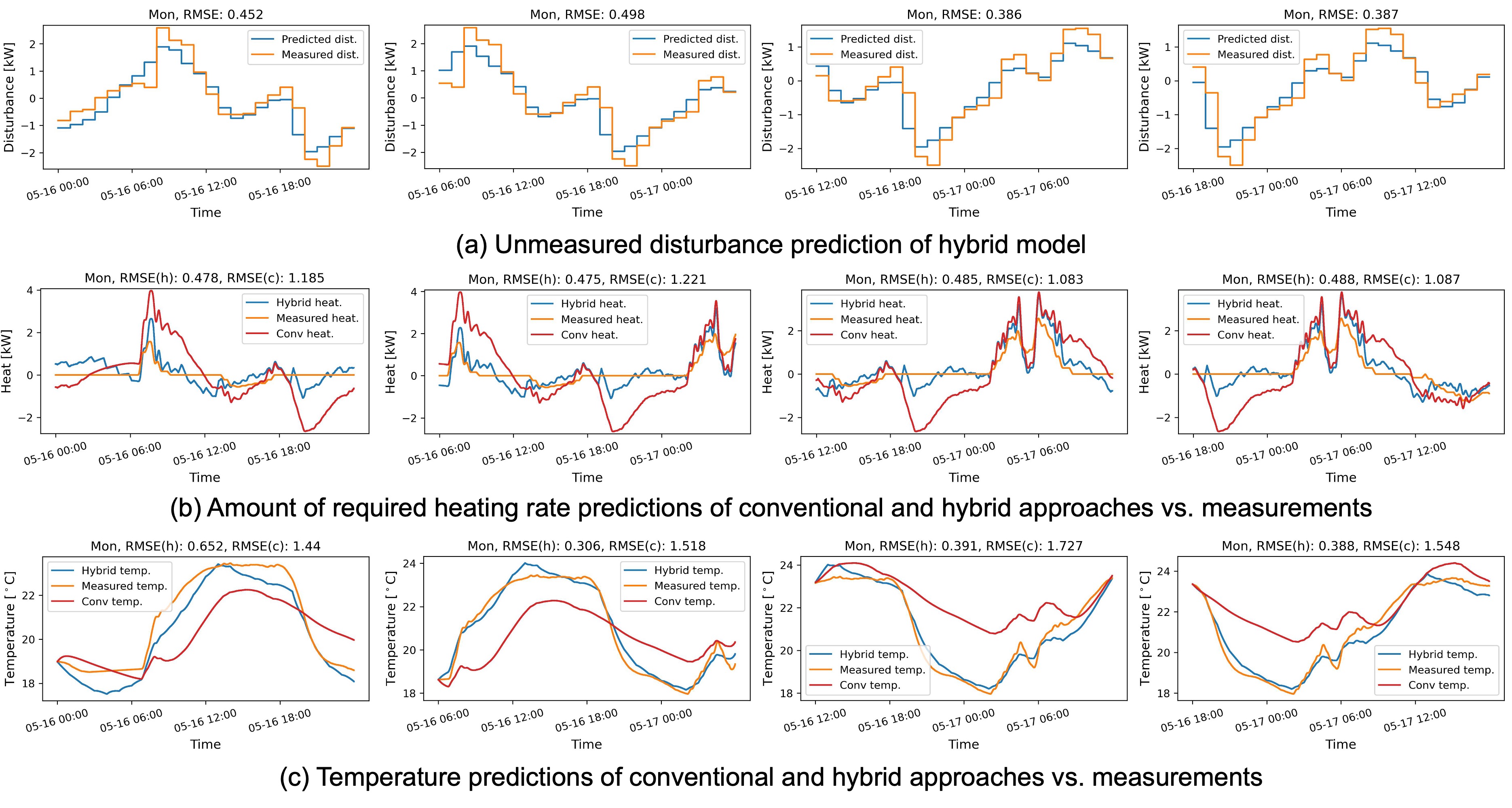} 
    \caption{Prediction performance of Hybrid approach vs. measured data and Conventional approach} 
    \label{fig:flexlab_prediction} 
\end{figure}

Figure~\ref{fig:flexlab_prediction} compares the prediction performance of the Hybrid approach against measured data and the Conventional approach. The data is visualized in 6-hour intervals to assess prediction accuracy at different times of day. Results show that the Hybrid model effectively predicts future unmeasured disturbances, achieving an average RMSE reduction of 1.3$^\circ$C in temperature prediction. In contrast, the absence of unmeasured disturbance prediction in the Conventional approach leads to increased inaccuracies in both temperature prediction and required heating rate, which can negatively affect the performance of predictive control applications.  

\section{Conclusion and limitations}
\label{sec:body5}

In this paper, we present a Hybrid modeling approach that enhances the long-term temperature and load prediction capabilities of a gray-box model. While gray-box models are widely accepted for predictive applications such as MPC for energy efficiency and decarbonization, their system identification becomes challenging under unmeasured disturbances such as occupancy, lighting, appliances, and infiltration/exfiltration loads. Since directly measuring these loads is costly and often impractical in real buildings, we propose a neural network–based model for unmeasured disturbance loads to address the inherent limitations of the gray-box approach.  

After comparing the performance and limitations of various system identification methods developed to handle unmeasured disturbances, we introduce the Hybrid modeling approach. To address the overfitting risk of neural network–based models, we develop a structured design and model selection process incorporating statistical tests. Using a calibrated building model, we generate two realistic datasets from different climates and use them to develop and evaluate the Hybrid models.  

The Hybrid approach achieves RMSE reductions of approximately $0.2$–$0.9^\circ$C and $0.3$–$2^\circ$C for one-day-ahead temperature predictions in mild (Berkeley, CA) and cold (Chicago, IL) climates, respectively. Furthermore, when applied to experimental data from an office-setting laboratory building, the Hybrid approach reduces RMSE by an average of $1.3^\circ$C compared to the Conventional approach.  

Despite these improvements, non-negligible prediction errors remain due to the stochastic nature of unmeasured disturbances, making the development of a perfect model impossible. In addition, when the gray-box model is inaccurate, the Hybrid model structure can be flawed if future control inputs are not incorporated, as illustrated in Fig.~\ref{fig:inaccurate}. While it is possible to include future control inputs in the Hybrid model, doing so introduces nonlinearity, which may impose computational burdens or degrade performance in predictive applications such as MPC.  

Moreover, true unmeasured disturbances may vary over time due to changes in building operations or occupancy profiles. Although real-time modeling with sequential updates \cite{Ham2022-mq} can address this issue, it remains advisable to recalibrate the model regularly or after significant operational changes (e.g., seasonal transitions or extended building closures).  

Overall, the Hybrid model offers superior predictive accuracy compared to the Conventional approach, particularly in compensating for inaccuracies in gray-box models affected by unmeasured disturbances. Since MPC makes optimal decisions at each control step, capturing the overall future thermal behavior of building operations is essential. Therefore, the Hybrid modeling approach can play a crucial role in enhancing energy efficiency and supporting decarbonization efforts by enabling long-term prediction applications such as load shifting and renewable energy integration.  

\section{Acknowledgements}
This work was supported by the Assistant Secretary for Energy Efficiency and Renewable Energy, Building Technologies Office, of the U.S. Department of Energy under Contract No. DE-AC02-05CH11231, by California Energy Commission through grant EPC-19-013, and by Korea Institute of Energy Technology Evaluation and Planning (KETEP) and the Ministry of Trade, Industry \& Energy (MOTIE) of the Republic of Korea (No. 20212020800120).

%\end{linenumbers}

 \bibliographystyle{elsarticle-num} 
 \bibliography{paperpile}





% \end{thebibliography}
\appendix
\section{Input features by model cases}

The following input features are assigned for model cases. Table \ref{tb:a1} shows the input features of MLP and CNN models, and those of RNN and LSTM are presented in Table \ref{tb:a2}

\begin{table}[!htbp]
\centering
\begin{tabular}{llll}
\hline
case   & pattern features & past $w$ & future $w$ \\ 
\hline
case01& 1 day&$\zeta_\text{ID}$,how,$T_\text{oa}$,$q_\text{sol,win}$&how,$T_\text{oa}$,$q_\text{sol,surface}$\\
case02 & 4 days & $\zeta_\text{ID}$ & $T_\text{oa}$,$q_\text{sol,win}$ \\
case03 & 4 days & $\zeta_\text{ID}$,how& how,$T_\text{oa}$,$q_\text{sol,win}$     \\
case04 & 4 days& $\zeta_\text{ID}$,weekday& weekday,$T_\text{oa}$,$q_\text{sol,win}$ \\
case05 & 4 days& $\zeta_\text{ID}$,how,$T_\text{oa}$,$q_\text{sol,win}$ & how,$T_\text{oa}$,$q_\text{sol,win}$     \\
case06 & 4 days& $\zeta_\text{ID}$,weekday,$T_\text{oa}$,$q_\text{sol,win}$& weekday,$T_\text{oa}$,$q_\text{sol,win}$ \\
case07 & 7 days& $\zeta_\text{ID}$& $T_\text{oa}$,$q_\text{sol,win}$           \\
case08 & 7 days& $\zeta_\text{ID}$,how& how,$T_\text{oa}$,$q_\text{sol,win}$     \\
case09 & 7 days& $\zeta_\text{ID}$,weekday& weekday,$T_\text{oa}$,$q_\text{sol,win}$ \\
case10 & 7 days& $\zeta_\text{ID}$,how,$T_\text{oa}$,$q_\text{sol,win}$& how,$T_\text{oa}$,$q_\text{sol,win}$\\
case11 & 7 days& $\zeta_\text{ID}$,weekday,$T_\text{oa}$,$q_\text{sol,win}$  & weekday,$T_\text{oa}$,$q_\text{sol,win}$ \\
case12 & 4 days& $\zeta_\text{ID}$,dow,$T_\text{oa}$,$q_\text{sol,win}$& dow,$T_\text{oa}$,$q_\text{sol,win}$     \\
case13 & 7 days& $\zeta_\text{ID}$,dow,$T_\text{oa}$,$q_\text{sol,win}$& dow,$T_\text{oa}$,$q_\text{sol,win}$     \\
case14 & 4 days& $\zeta_\text{ID}$,$T_\text{oa}$,$q_\text{sol,win}$& $T_\text{oa}$,$q_\text{sol,win}$           \\
case15 & 7 days& $\zeta_\text{ID}$,$T_\text{oa}$,$q_\text{sol,win}$& $T_\text{oa}$,$q_\text{sol,win}$           \\
case16 & 4 days& $\zeta_\text{ID}$,weekday,$T_\text{oa}$,$q_\text{sol,win}$,$i_\text{heat}$,$i_\text{cool}$ & dow,$T_\text{oa}$,$q_\text{sol,win}$     \\
case17 & 7 days& $\zeta_\text{ID}$,weekday,$T_\text{oa}$,$q_\text{sol,win}$,$i_\text{heat}$,$i_\text{cool}$ & dow,$T_\text{oa}$,$q_\text{sol,win}$     \\
case18 & 4 days& $\zeta_\text{ID}$,dow,$T_\text{oa}$,$q_\text{sol,win}$,$i_\text{heat}$,$i_\text{cool}$     & dow,$T_\text{oa}$,$q_\text{sol,win}$     \\
case19 & 7 days& $\zeta_\text{ID}$,dow,$T_\text{oa}$,$q_\text{sol,win}$,$i_\text{heat}$,$i_\text{cool}$     & dow,$T_\text{oa}$,$q_\text{sol,win}$     \\
case20 & 4 days& $\zeta_\text{ID}$,how,$T_\text{oa}$,$q_\text{sol,win}$,$i_\text{heat}$,$i_\text{cool}$     & dow,$T_\text{oa}$,$q_\text{sol,win}$     \\
case21 & 7 days& $\zeta_\text{ID}$,how,$T_\text{oa}$,$q_\text{sol,win}$,$i_\text{heat}$,$i_\text{cool}$     & dow,$T_\text{oa}$,$q_\text{sol,win}$     \\ \hline
\end{tabular}
\caption{Input features of MLP and CNN models for model cases}
\label{tb:a1}
\end{table}

\begin{table}[!htbp]
\centering
\begin{tabular}{llll}
\hline
case   & pattern feature & past $w$                                        & future $w$                                 \\ \hline
case01 & 1 day& $\zeta_\text{ID}$,how,$T_\text{oa}$,$q_\text{sol,win}$     & how,$T_\text{oa}$,$q_\text{sol,win}$     \\
case02 & 1 day& $\zeta_\text{ID}$,how                               & how                               \\
case03 & 1 day& $\zeta_\text{ID}$,$T_\text{oa}$,$q_\text{sol,win}$           & $T_\text{oa}$,$q_\text{sol,win}$           \\
case04 & 1 day& $\zeta_\text{ID}$,hod,$T_\text{oa}$,$q_\text{sol,win}$     & hod,$T_\text{oa}$,$q_\text{sol,win}$     \\
case05 & 1 day& $\zeta_\text{ID}$,dow,$T_\text{oa}$,$q_\text{sol,win}$     & dow,$T_\text{oa}$,$q_\text{sol,win}$     \\
case06 & 1 day& $\zeta_\text{ID}$,weekday,$T_\text{oa}$,$q_\text{sol,win}$ & weekday,$T_\text{oa}$,$q_\text{sol,win}$ \\
case07 & 2 days& $\zeta_\text{ID}$,how,$T_\text{oa}$,$q_\text{sol,win}$     & how,$T_\text{oa}$,$q_\text{sol,win}$     \\
case08 & 4 days& $\zeta_\text{ID}$,how,$T_\text{oa}$,$q_\text{sol,win}$     & how,$T_\text{oa}$,$q_\text{sol,win}$     \\
case09 & 7 days& $\zeta_\text{ID}$,how,$T_\text{oa}$,$q_\text{sol,win}$     & how,$T_\text{oa}$,$q_\text{sol,win}$     \\
case10 & 4 days& $\zeta_\text{ID}$,$T_\text{oa}$,$q_\text{sol,win}$           & $T_\text{oa}$,$q_\text{sol,win}$           \\
case11 & 7 days& $\zeta_\text{ID}$,$T_\text{oa}$,$q_\text{sol,win}$           & $T_\text{oa}$,$q_\text{sol,win}$           \\
case12 & 2 days& $\zeta_\text{ID}$,weekday,$T_\text{oa}$,$q_\text{sol,win}$ & weekday,$T_\text{oa}$,$q_\text{sol,win}$ \\
case13 & 4 days& $\zeta_\text{ID}$,weekday,$T_\text{oa}$,$q_\text{sol,win}$ & weekday,$T_\text{oa}$,$q_\text{sol,win}$ \\
case14 & 7 days& $\zeta_\text{ID}$,weekday,$T_\text{oa}$,$q_\text{sol,win}$ & weekday,$T_\text{oa}$,$q_\text{sol,win}$ \\
case15 & 2 days& $\zeta_\text{ID}$,dow,$T_\text{oa}$,$q_\text{sol,win}$     & dow,$T_\text{oa}$,$q_\text{sol,win}$     \\
case16 & 4 days& $\zeta_\text{ID}$,dow,$T_\text{oa}$,$q_\text{sol,win}$     & dow,$T_\text{oa}$,$q_\text{sol,win}$     \\
case17 & 7 days& $\zeta_\text{ID}$,dow,$T_\text{oa}$,$q_\text{sol,win}$     & dow,$T_\text{oa}$,$q_\text{sol,win}$     \\ \hline
\end{tabular}
\caption{Input features of RNN and LSTM models for model cases}
\label{tb:a2}
\end{table}

\end{document}